\documentclass[%
 reprint,
 amsmath,amssymb,
 aps
]{revtex4-2}

\usepackage{graphicx}
\usepackage{dcolumn}
\usepackage{bm}
\usepackage{array}
\usepackage[T1]{fontenc}
\usepackage{array}
\usepackage[normalem]{ulem}
\usepackage{physics}
\usepackage[latin1,utf8]{inputenc}
\usepackage[OT2,OT1]{fontenc}
\usepackage{color}

\begin{document}

\title{The effect of screening mechanisms on black hole binary inspiral waveforms}

\author{Cyril Renevey}
\email{renevey@lpsc.in2p3.fr}
\affiliation{Laboratoire de Physique Subatomique et Cosmologie, \\ 53 Avenue des Martyrs, 38000 Grenoble, France}
\author{Ryan McManus}
\email{ppzrm@exmail.nottingham.ac.uk}
\affiliation{School of Physics and Astronomy, University of Nottingham, \\
University Park, Nottingham NG7 2RD, United Kingdom}
\author{Charles Dalang}
\email{charles.dalang@unige.ch}
\affiliation{D\'epartement de Physique Th\'eorique, Universit\'e de Gen\`eve, \\ 24 quai Ernest Ansermet, 1211 Gen\`eve 4, Switzerland}
\author{Lucas Lombriser}
\email{lucas.lombriser@unige.ch}
\affiliation{D\'epartement de Physique Th\'eorique, Universit\'e de Gen\`eve, \\ 24 quai Ernest Ansermet, 1211 Gen\`eve 4, Switzerland}

\date{\today}

\definecolor{darkgreen}{rgb}{0,0.5,0}
\newcommand{\edit}[1]{{\color{darkgreen}{#1}}}
\newcommand{\location}[1]{{\color{red}{#1}}}
\newcommand{\lucas}[1]{{\color{red}{[LL:~#1]}}}
\definecolor{ao(english)}{rgb}{0.0, 0.5, 0.0}
\newcommand{\done}[1]{\textcolor{ao(english)}{#1}}
\newcommand{\tabspace}{\vphantom{$\frac{\frac{x^2}{x}}{\frac{x}{x}}$}}
\newcommand{\ryan}[1]{{\color{red}{#1}}}
\newcommand{\sums}{\sum\limits}
\newcommand{\I}{{(i)}}
\newcommand{\J}{{(j)}}
\newcommand{\munu}{{\mu\nu}}
\newcommand{\bigcups}{\bigcup\limits}
\newcommand{\0}{{(0)}}
\newcommand{\1}{{(1)}}
\newcommand{\2}{{(2)}}
\newcommand{\3}{{(3)}}
\newcommand{\q}{{(q)}}
\newcommand{\opn}{\mathcal{O}_{PN}}
\newcommand{\oa}{\mathcal{O}_{\alpha}}
\newcommand{\charles}[1]{\textcolor{blue}{[Charlie: #1]}}
\newcommand{\cyril}[1]{\textcolor{red}{Cyril: #1}}
\newcommand{\RM}[1]{\textcolor{red}{Ryan: #1}}
\newcommand{\e}[1]{_{\rm #1}}

\def \bs {\boldsymbol}
\def \be {\begin{equation}}
\def \ee {\end{equation}}
\def \dd {\mathrm{d}}
\def \t {\tilde}
\def \p {\partial}
\def \l {\left}
\def \r {\right}
\def \te {\tensor} 

\begin{abstract}
Scalar-tensor theories leaving  significant modifications of gravity at cosmological scales rely on screening mechanisms to recover General Relativity (GR) in high-density regions and pass stringent tests with astrophysical objects.
Much focus has been placed on the signatures of such modifications of gravity on the propagation of gravitational waves (GWs) through
cosmological distances while typically assuming their emission from fully screened regions with the wave generation strictly abiding by GR.
Here, we closely analyse the impact of screening mechanisms on the inspiral GW waveforms from compact sources by employing a scaling method that enables a post-Newtonian (PN) expansion in screened regimes. Particularly, we derive the leading-order corrections to a fully screened emission to first PN order in the near zone and we also compute the modifications in the unscreened radiation zone to second PN order.
For a concrete example, we apply our results to a cubic Galileon model.
The resulting GW amplitude from a binary black hole inspiral deviates from its GR counterpart at most by one part in $10^{2}$ for the modifications in the radiation zone and at most one part in $10^{11}$ due to next-order corrections to the fully screened near zone.
We expect such modifications to be undetectable by the current generation of GW detectors, but the deviation is not so small as to remain undetectable in future experiments. 
\end{abstract}

\maketitle

\onecolumngrid
\section{Introduction}
\label{sec:intro}

General relativity (GR) and quantum field theory constitute the cornerstones of theoretical physics and allow for precise predictions which have been successfully tested experimentally \cite{Will:2014kxa,Chatrchyan:2012ufa,Aad:2012tfa}. In the context of cosmology, the observational viability of GR invokes a Universe dominated by dark fluids~\cite{Aghanim:2018eyx}: a dark energy, explaining the late-time accelerated expansion of our Universe, and cold dark matter, describing the cosmic microwave background anisotropies, galaxy clustering and lensing as well as galactic rotation curves.
For a dark energy in form of a cosmological constant this composition constitutes the concordance cosmology, the $\Lambda$-Cold-Dark-Matter ($\Lambda$CDM) model.
Despite its observational success, a number of smaller and larger tensions have recently arisen that remain unexplained to date~\cite{Riess:2020fzl,Wong:2019kwg,DiValentino:2019qzk,Handley:2019tkm,Asgari:2020wuj,Lombriser:2019ahl,Bose:2020cjb,DiValentino:2021izs,Perivolaropoulos:2021jda}.
Perhaps an even larger enigma is posed by our lack of understanding of the cosmological constant.
If attributed to the quantum fluctuations of vacuum energy,
theoretical estimates for its magnitude from quantum field theory are off by more than $50$ orders of magnitude~\cite{Weinberg:1988cp,Martin:2012bt,Kaloper:2013zca,Lombriser:2019jia}.
One of the leading hypotheses for this discrepancy is the existence of new symmetries, 
that are thought to cancel vacuum fluctuations.
This, however, leaves the need for an explanation of the small residual cosmological constant driving the observed late-time accelerated cosmic expansion.

In this context,
a range of alternative explanations for cosmic acceleration to this residual cosmological constant have (re-)emerged over the past two decades, which involve adding new degrees of freedom to the gravitational action~\cite{Brans:1961sx,Horndeski:1974wa,deRham:2010kj,Heisenberg:2014rta,Joyce:2016vqv,Heisenberg:2018vsk}.
Of these alternative theories of gravity, the Horndeski action~\cite{Horndeski:1974wa} constitutes the most general class of Lorentz-invariant four-dimensional scalar-tensor theories that lead to second-order equations of motion, which turns out to be sufficient~\cite{Ostrogradsky1850} but not necessary~\cite{deRham:2011qq,Gleyzes:2014dya,Jana:2020vov} to avoid Ostrogradski instabilities.
A significant parameter space of this class of models has traditionally been explored as an explanation of cosmic acceleration.
Often, a significant modification of gravity on cosmological scales was invoked as the driver of cosmic acceleration, while so-called \textit{screening mechanisms} would allow a recovery of GR in high-density environments or at short distances such as in galaxies or in the Solar System, where tight constraints on deviations from GR have been established~\cite{Stairs:2003eg,Will:2014kxa,Manchester:2015mda,Archibald:2018oxs,Will:2018bme,Renevey:2019jrm,Baker:2019gxo}.
However, this concept has become severely challenged by the luminal propagation of gravitational waves (GWs)~\cite{Lombriser:2016yzn,Monitor:2017mdv}, and cosmic acceleration may be limited to the dark energy aspects of a Horndeski field rather than its direct dynamical change of gravity.
Nevertheless, dark energy fields may be accompanied by cosmologically significant universal interactions with the matter fields, which requires screening on small scales and for which screening effects can provide distinctive observational signatures~\cite{Baker:2019gxo}.

Prominent screening mechanisms include the chameleon~\cite{Khoury:2003rn}, k-mouflage~\cite{Babichev:2009ee} and Vainshtein~\cite{Vainshtein:1972sx} mechanisms, which rely on non-linear terms in the equations of motion to recover GR in their respective screened regimes.
Consequently, the non-linearity renders post-Newtonian approaches difficult to implement due to the linearisation of the relevant field equations. 
However, it was shown that this problem can be circumvented with the employment of the so-called \textit{scaling method}~\cite{McManus:2016kxu,Lombriser:2016zfz,McManus:2017itv,Renevey:2020tvr} that enables an expansion in screened regions.
More specifically, the method can be used to both identify the leading corrections from the scalar field to the  metric field equations in the screened small-scale regions or conversely of the unscreened theory on large scales.

The first direct detection of GWs in 2015~\cite{Abbott:2016blz} has opened a new observational window on gravity.
To exploit this new wealth of data, a range of analytical and numerical techniques have been developed to study the inspiral-merger-ringdown of two heavy compact objects \cite{1983grr..proc....1T,Maggiore:1900zz, Maggiore:2018sht}.
Among them is the post-Newtonian (PN) formalism for the inspiral phase, where a term in the waveform is said to be of PN order $n$ if it is of order $\mathcal{O}((v/c)^{2n})$, abbreviated $\mathcal{O}(\epsilon^n)$, where $v$ is the speed of any of the two objects. 
In GR, the waveform from an inspiral system has already been determined to 5PN~\cite{PhysRevLett.122.241605,BLUMLEIN2020135100}, with multiple complimentary methods of calculation. 
In the case of Brans-Dicke theories~\cite{Brans:1961sx}, the calculations have been performed up to 2PN by Lang~\cite{Lang:2013fna} following the method developed by Will and Wiseman~\cite{Will:1996zj}.
In this study, we will use these two references extensively and refer to them from now on as Lang14 and WW96, respectively.
In the context of modified gravity, numerous studies have explored how propagation effects modify the amplitude or polarization of the GW~\cite{Lombriser:2015sxa,Nishizawa:2017nef,Belgacem:2017ihm,Amendola:2017ovw,Lagos:2019kds,Belgacem:2019pkk,Dalang:2019fma,Dalang:2019rke,Garoffolo:2019mna,Mastrogiovanni:2020gua,Baker:2020apq,Dalang:2020eaj,Fanizza:2020hat,Mukherjee:2020mha}.
Predominantly, these studies assume that the generation of the waveform can be modelled according to GR, invoking the operation of a screening mechanism at the source as motivation for that.

\begin{figure}
  \centering
  \includegraphics[width=0.5\linewidth]{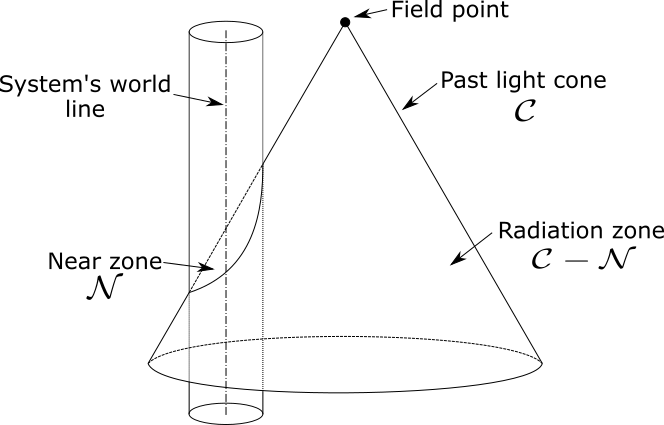}
  \caption{
  Adapted from \cite{Will:1996zj}. The past light cone $\mathcal{C}$ of a field point far from the binary system. 
  The near zone, $\mathcal{N}$, is the intersection of the past light cone and the world tube of the system's world line with radius $\mathcal{R}$. 
  The radiation zone, $\mathcal{C}-\mathcal{N}$, is the remainder of the light cone.
  We assume that the metric field equations take on their screened form within $\mathcal{N}$ whereas they assume their unscreened form in $\mathcal{C}-\mathcal{N}$. When calculating the waveform at the field point, we treat the contribution from the near zone and radiation zone separately.
  }
  \label{fig:pastLightCone}
\end{figure}

In this paper, we compute the leading-order corrections to the GW waveform expected for the three aforementioned
screening mechanisms\footnote{Note that numerical attempts to describe GW radiation for theories exhibiting the Vainshtein mechanism, have been conducted in Ref.~\cite{deRham:2012fw,deRham:2012fg,Dar:2018dra}, but its inspiral waveform and that of other screening mechanisms have not been described previously.}, including both effects on propagation and the leading correction to the screened GW generation.
To this end, we follow the method of WW96 and cast the field equations in their relaxed form, i.e., as a flat-space wave equation sourced by quadratic self-interactions and the stress-energy tensor. 
In this form, the solutions are retarded Green's integrals, for which the integrands depend on the GW itself, the quantity we are trying to solve for. 
The problem can be solved perturbatively by first neglecting the self-interactions, providing an approximate solution $h^{\mu\nu}_{(1)}$, where the only source is the energy-momentum tensor, and introducing those in the retarded integrals to obtain a new, more accurate solution $h^{\mu\nu}_{(2)}$. 
In practice, the integrals are split between the so-called \textit{near} and \textit{radiation} zones, which are depicted in Fig.~\ref{fig:pastLightCone}.
These are complementary spacelike volumes with their boundary carrying no physical meaning. 
In GR, the volume separation is introduced to cancel apparently divergent surface integrals between the near and radiation zones. 
In this work, we use this splitting in a physically motivated sense such that the near zone corresponds to a screened region, where nevertheless small deviations with respect to GR may arise at the next-to-leading order, and where an unscreened scalar-tensor modification applies in the radiation zone.
Besides determining the overall modified waveform, this strategy 
enables us to compute the leading-order corrections on the generation of the waveform in modified gravity theories, which so far has typically been assumed fully screened and abiding exactly by GR.

The paper is organised as follows. 
In Sec.~\ref{sec:modified_equations_with_screening}, we derive the equations of motion for a subset of Horndeski theories that describe tensorial GWs propagating at light speed and which have sufficient freedom to allow for one of the three types of screening mechanisms. 
We express those in their relaxed form, briefly review the scaling method and explain how it applies to the retarded integrals of the Green's function.
In Sec.~\ref{sec:waveform_fully_screened}, we combine the results of Lang14 and WW96, assuming first the fully screened scenario, where the scalar field has no effect in the near zone and GR applies exactly.
We compute the leading terms that arise from Brans-Dicke corrections in the radiation zone to 2PN.
In Sec.~\ref{sec:corrections_to_waveform_from_scalar_field}, we then compute the leading correction to the fully screened case from the scalar field in the near zone by employing the scaling method.
Further, we calculate the gravitational waveform up to 1PN beyond the quadrupole formula.
In Sec.~\ref{sec:reduced_case_binary_black_hole}, we compute those corrections for a black hole binary system and apply it to an example of Galileon cosmology.
Finally, Sec.~\ref{sec:Conclusions} is dedicated to our conclusions and Appendix~\ref{app:CalculatingTheGravitationalWaveform} provides further details on our calculation of the gravitational waveform from a binary system in the case of GR.

Throughout the paper, we adopt the signature $(-,+,+,+)$ for the metric. Greek indices run from $0$ to $3$ and Latin indices from $1$ to $3$. A comma indicates a partial derivative, $A_{,\mu}\equiv \partial_\mu A$, while a semicolon denotes a covariant derivative associated with the Levi-Civita connection, $A_{\mu;\nu}\equiv \nabla_\nu A_\mu$. Bold symbols $\bs{x}$ represent Euclidean three-dimensional vectors with Euclidean norm $|\bs{x}|$. Symmetrization of indices follow the notation $T_{(\mu_1\dots\mu_n)}\equiv \frac{1}{n!}\sums_{i\in S_n}T_{\sigma_i(\mu_1\dots\mu_n)}$, where $\{\sigma_i \}_{i\in S_n}$ is the set of all permutations of $\{\mu_1,\dots,\mu_n\}$.  Units are such that $c=\hbar=1$.

\section{Modified field equations in the presence of screening}\label{sec:modified_equations_with_screening}

We will consider scalar-tensor theories that can be described by the following action,
\begin{align}
    S[g,\phi]=\frac{1}{16\pi G}\int\sqrt{-g}\left(\phi R+2\frac{\omega(\phi)}{\phi}X+\alpha \delta_CV(\phi)+\alpha\delta_K\Gamma_2(\phi)X^2+\alpha\delta_V\Gamma_3(\phi)X\Box\phi\right)\,,\label{eq:action_screened_horndeski}
\end{align}
where $R$ is the Ricci scalar of the metric $g_{\mu\nu}$ with determinant $g \equiv \det g_{\mu\nu}$.
$V,\omega,\Gamma_2,\Gamma_3$ are arbitrary functions of the scalar field $\phi$, $X=- g^{\mu\nu}\partial_\mu \phi\partial_\nu\phi/2$ is its kinetic term and $\Box \equiv g^{\mu\nu}\nabla_\mu \nabla_\nu$.
The action~\eqref{eq:action_screened_horndeski} corresponds to a reduced set of Horndeski theories~\cite{Horndeski:1974wa} with luminal propagation of tensor gravitational waves~\cite{McManus:2016kxu}, satisfying arrival time constraints~\cite{Lombriser:2015sxa} from GW170817~\cite{Monitor:2017mdv}. The functions $V(\phi)$, $\Gamma_2(\phi)$ and $\Gamma_3(\phi)$ allow for large field value as well as first and second derivative screening, respectively~\cite{Joyce:2016vqv}.
While Eq.~\eqref{eq:action_screened_horndeski} in principle admits a superposition of screening mechanisms, we will assume for simplicity that only one of the screening effects is operating at a time. 
The parameters $\delta_C$, $\delta_K$ and $\delta_V$ are unity when implementing the chameleon \cite{Khoury:2003rn}, k-mouflage \cite{Babichev:2009ee} and Vainshtein \cite{Vainshtein:1972sx} screening mechanisms, respectively, and vanish otherwise.
The \emph{scaling parameter} $\alpha$ will be utilised to identify the most relevant terms beyond GR in the screened regime. The field equations derived from the action~\eqref{eq:action_screened_horndeski} are of the form
\begin{align}
\label{eq:genericMetricFieldEquation}
\phi G_{\mu\nu} &= 8 \pi G T_{\mu\nu} + \frac{\omega(\phi)}{\phi}(\phi_{,\mu}\phi_{,\nu} - \frac{1}{2} g_{\mu\nu}\phi_{,\lambda}\phi^{,\lambda}) + (\nabla_\mu\nabla_\nu\phi - g_{\mu\nu}\Box \phi) + \alpha A_{\mu\nu}[g,\phi] \,, \\
\label{eq:genericScalarFieldEquation}
\Box \phi &= \frac{1}{3+2\omega(\phi)}\big(8 \pi G T- 16\pi G \phi \pdv{T}{\phi} - \dv{\omega}{\phi}\,\phi_{,\lambda}\phi^{,\lambda}+\alpha A^\lambda_{\ \,\lambda}[g,\phi]\big) + \alpha B[g,\phi]\,,
\end{align}
where $G_{\mu\nu}$ is the Einstein tensor, $T_{\mu\nu}$ is the stress-energy (or energy-momentum) tensor of the matter fields and $T\equiv g^{\mu\nu} T_{\mu\nu}$ denotes its trace.
The second term on the right-hand side of the scalar field equation of motion appears if we assume that the energy-momentum tensor effectively depends on the scalar field~\cite{Eardley:1975aaa}.
The additional two functionals $A_\munu[g, \phi]$ and $B[g, \phi]$ include the contributions of the functions $V,\,\Gamma_2$ and $\Gamma_3$, namely
\begin{align}
    A_\munu=&\ \delta_C\,\frac{1}{2}\,g_\munu V(\phi)+\delta_K\,\Gamma_2(\phi)\left(\frac{1}{2}\, g_\munu X^2+X \nabla_\mu\phi\nabla_\nu\phi\right)+\delta_V\, \pdv{\Gamma_3}{\phi}X\nabla_\mu\nabla_\nu\phi
    \nonumber\\
    &+ \delta_V\, \Gamma_3(\phi)\left(\frac{1}{2}\,g_\munu\nabla^\alpha\phi\nabla_\beta\nabla_\alpha\phi\nabla^\beta\phi+\Box\phi\nabla_\mu\phi\nabla_\nu\phi- \nabla^\alpha\phi\nabla_\alpha\nabla_{(\mu}\phi\nabla_{\nu)}\phi\right)\, ,
    \\
    B=&\ -\delta_C\, \pdv{V}{\phi}-\delta_K\,\Gamma_2(\phi)\left(2\,X\,\Box\phi-2\nabla_\alpha\phi\nabla_\beta\phi\nabla^\alpha\nabla^\beta\phi\right)+3\,\delta_K\,\pdv{\Gamma_2}{\phi}X^2
    \nonumber\\
    & -\delta_V\, \Gamma_3(\phi)\left((\Box\phi)^2-\nabla_\alpha\nabla_\beta\phi\nabla^\alpha\nabla^\beta\phi\right)+2\, \delta_V\, \left(\pdv{\Gamma_3}{\phi}\nabla_\alpha\phi\nabla_\beta\phi\nabla^\alpha\nabla^\beta\phi+\pdv[2]{\Gamma_3}{\phi}X^2\right)\, .
\end{align} 
The role of screening mechanisms is to
suppress
the gravitational effect of the scalar field in high-density regions. Therefore, the functionals $A_\munu$ and $B$ are expected to have a significant effect in the near zone, so as to recover GR, whereas they are expected to vanish far away from the source ($\alpha\rightarrow0$).

For the matter source, we consider a collection of point particles, labelled with capital indices, which are described by the following energy-momentum tensor at the field point $x^\gamma=(t,\bs{x})$,
\begin{equation}
\label{eq:MatterDistribution}
T^{\alpha\beta}(x^\gamma) = \sum_A m_A(\phi) (-g)^{-1/2} \frac{u^\alpha_A u^\beta_A}{u^0_A} \delta^3(\bs{x}- \bs{x}_A(t))\,,
\end{equation}
where $u^\alpha_A$ is the four-velocity of body $A$, $\bs{x}_A(t)$ is its time-dependent spatial position and $m_A$ its gravitational mass including the gravitational binding energy of the $A$-th body.
An unfortunate consequence of modeling the matter source as point particles is that we neglect tidal effects, but see \cite{Le_Tiec_2021} for recent work on tidal deformability of black holes.

Note that the masses can depend on the scalar field, even in the Jordan frame we have adopted~\cite{Eardley:1975aaa}. 
With the assumption of massive point-like objects, we neglect non-trivial effects of $\phi$ on the matter content, which violate the strong equivalence principle. For Brans-Dicke theories, one needs to take into account the dependence of the gravitational constant on the scalar field since it affects the total gravitational energy. As suggested in Ref.~\cite{Eardley:1975aaa}, when considering the inertial mass of objects in the Jordan frame, the masses of compact object are replaced with functions of the scalar field where the total mass is given by
\begin{equation}
m\equiv \sum_A m_A(\phi)\,.
\end{equation}
The inertial mass can then be expanded about the local background value of the scalar field $\phi_0$,
\begin{equation}
    m_A(\phi) \approx m_A(\phi_0)(1 + s_A (\phi-\phi_0) / \phi_0 )\,,
\end{equation}
where
\begin{equation}
    s_A = - \frac{\partial \ln m_A}{\partial \ln G_N} \bigg |_{\phi_0} = \frac{\partial \ln m_A}{\partial \ln \phi} \bigg |_{\phi_0}
\end{equation}
is the sensitivity of compact objects in scalar-tensor theories and in particular, $s_A=1/2$ for black holes. However, for theories exhibiting Vainshtein or k-mouflage mechanisms the scalar field equation in the screened region effectively becomes a total derivative. One can then use the generalized Gauss theorem to stipulate that the scalar field cannot depend on the composition of the object, but only on its total mass~\cite{Chow:2009fm}. Therefore, we only need to take the sensitivity into account in the case of the chameleon mechanism and the inertial mass becomes
\begin{equation}
    m_A(\phi) \approx m_A(1 + \delta_C s_A (\phi-\phi_0) / \phi_0 )\, ,
\end{equation}
where we renamed $m_A(\phi_0)\rightarrow m_A$ for simplicity.

With the equations of motion at hand and an explicit matter source specified, we proceed to compute the waveform at a generic spacetime point far away from the source, in the so-called \emph{radiation zone}. To this end, we make use of the field equations~\eqref{eq:genericMetricFieldEquation} and \eqref{eq:genericScalarFieldEquation} in their relaxed forms.
These are obtained through a series of field redefinition. As the scalar field couples non-minimally to the metric, their derivation is different from that in GR and we follow Ref.~\cite{Mirshekari:2013vb} to find them.
The scalar field is redefined as 
\begin{equation}
\varphi(t,\bs{x}) \equiv \frac{\phi(t,\bs{x})}{\phi_0}\,,
\end{equation}
We assume it to be static for the purpose of this calculation as the time scales over which the background value varies is much larger than the orbital time scales of the system. Note that a change in the scalar field value between the spacetime positions of the source and observer may be detected as an effective modification of the luminosity distance~\cite{Dalang:2019fma, Dalang:2019rke}. 
We define the following symmetric rank-two tensor, which describes the deviation from flat space~\cite{Misner:1974qy}
\begin{equation}
\label{eq:gravPotential}
h^{\mu\nu} = \eta^{\mu\nu} - \varphi \sqrt{-g} g^{\mu\nu}\,.
\end{equation}
Note that we have not yet assumed that $h^{\mu\nu}$ is small. Next, we impose the flat-space harmonic gauge on $h^{\mu\nu}$,
\begin{equation}
h^{\mu\nu}{}_{,\nu} =0\,.
\end{equation}
The purpose for the field redefinition~\eqref{eq:gravPotential} is to rewrite the metric field equation~\eqref{eq:genericMetricFieldEquation} as a flat-space wave equation, 
\begin{equation}
\label{eq:screenedMetricRadiation}
\Box_\eta h^{\mu\nu} = - 16 \pi G \tau^{\mu\nu}-2\alpha(-g)A^\munu\frac{\phi}{\phi_0} \equiv S^{\mu\nu}\,,
\end{equation}
where $\Box_\eta \equiv \eta^{\mu\nu}\p_\mu \p_\nu$ and
\begin{equation}
\label{eq:GWSource}
\tau^{\mu\nu} \equiv (-g)\varphi T^{\mu\nu} + \frac{1}{16 \pi G}(\Lambda^{\mu\nu} + \Lambda_s^{\mu\nu})\,.
\end{equation}
One can think of $\Lambda^{\mu\nu}$ and $\Lambda_s^{\mu\nu}$ as the stress-energy tensors held in the gravitational wave and scalar field, respectively.
These in turn are given by
\begin{equation}
\label{eq:GravitationalStressEnergy}
\Lambda^{\mu\nu} \equiv 16 \pi G(-g) t^{\mu\nu}_{LL} + h^{\mu\alpha}{}_{,\beta}h^{\nu\beta}{}_{,\alpha} - h^{\alpha\beta}h^{\mu\nu}{}_{,\alpha\beta}\,,
\end{equation}
where 
\begin{align}
(-g) t^{\mu\nu}_{LL} &\equiv \frac{1}{16\pi G}\big[ 
g_{\lambda\alpha} g^{\beta \rho} h^{\mu\lambda}{}_{,\beta}h^{\nu\alpha}{}_{,\rho} 
+ \frac{1}{2}g_{\lambda\alpha}g^{\mu \nu}h^{\lambda\beta}{}_{,\rho}h^{\rho\alpha}{}_{,\beta}
-2g_{\alpha\beta}g^{\lambda(\mu}h^{\nu)\beta}{}_{,\rho}h^{\rho\alpha}{}_{,\lambda} \nonumber \\
&+\frac{1}{8}(2g^{\mu\lambda}g^{\nu\alpha} - g^{\mu\nu}g^{\lambda\alpha})(2 g _{\beta\rho}g_{\sigma\tau} - g_{\rho\sigma}g_{\beta\tau}) h^{\beta\tau}{}_{,\lambda}h^{\rho\sigma}{}_{,\alpha}
\big]
\end{align}
is the Landau-Lifshitz tensor and 
\begin{align}
\label{eq:LambdaS}
\Lambda_s^{\mu\nu} \equiv \frac{3+2\omega}{\varphi^2}\varphi_{,\alpha}\varphi_{,\beta}\left( (\eta^{\mu\alpha}- h^{\mu\alpha}) ( \eta^{\nu\beta} - h^{\nu\beta}) - \frac{1}{2}(\eta^{\mu\nu} - h^{\mu\nu}) (\eta^{\alpha\beta} - h^{\alpha\beta}) \right) \, .
\end{align}
The scalar field satisfies the equation 
\begin{equation}
\label{eq:screenedScalarRadiation}
\Box_\eta \varphi = - 8 \pi G \tau_s +\alpha B \equiv S \,,
\end{equation}
where the source of the scalar field equation is given by
\begin{align}
\label{eq:tauScalar}
\tau_s &= -\frac{1}{3+2\omega}\sqrt{-g}\varphi \left(T - 2 \varphi \frac{\partial T}{\partial \phi}\right)- \frac{1}{8\pi G}h^{\alpha\beta}\varphi_{,\alpha\beta}\nonumber \\
&+ \frac{1}{16\pi G}\frac{\dd}{\dd \varphi}\left[\ln\left(\frac{3+2\omega}{\varphi^2}\right)\right]\varphi_{,\alpha}\varphi_{,\beta}(\eta^{\alpha\beta} - h^{\alpha\beta})\,.
\end{align}

One may worry that second-order derivatives of the metric and scalar field appear within both $A^{\mu\nu}$ and $B$ in Eqs.~\eqref{eq:screenedMetricRadiation} and \eqref{eq:screenedScalarRadiation}.
However, the $\alpha$ prefactor in both terms ensures that the system in the screened and unscreened regime is appropriately diagonalised with respect to the second derivatives, with those appearing within $A^{\mu\nu}$ and $B$ acting as source terms for the next order in $\alpha$ corrections to the metric and scalar field.

Since the field equations take the form of a flat-space wave equation, one can express them as retarded integrals with the appropriate Green's function, namely
\begin{align}
\label{eq:hRetIntergral}
h^{\mu\nu}(t,\bs{x}) =&  -\int _\mathcal{C} \frac{S^{\mu\nu}(t-|\bs{x}-\bs{x'}|, \bs{x}')}{4\pi |\bs{x}-\bs{x'}|}\dd^3\bs{x'} \,, \\
\label{eq:PhiRetIntergral}
\varphi(t,\bs{x}) =& -\int _\mathcal{C} \frac{S(t-|\bs{x}-\bs{x'}|, \bs{x}')}{4\pi |\bs{x}-\bs{x}'|} \dd^3\bs{x'} \,,
\end{align}
where $\mathcal{C}$ is the past light cone of the field point $x^\mu$.
Note that the source terms $S$ depends both on $\varphi$ and $h^{\mu\nu}$, which makes Eqs.~\eqref{eq:hRetIntergral} and \eqref{eq:PhiRetIntergral} coupled integral equations for $\varphi$ and $h^{\mu\nu}$.

As a first step, in this work, we
specify to cases where the contribution of the scalar wave can be neglected. We leave the computations for the more general case to future work.
Depending on screening mechanism, the effect of an incoming scalar wave on our detectors may or may not be attenuated by screening in the Solar System. For example, in the chameleon screening mechanism, the suppression of scalar waves depends on the ratio between the effective mass of the scalar field and the GW frequency~\cite{Katsuragawa:2019uto}. However, not all mechanisms screen incoming scalar waves. For example, the Vainshtein mechanism does not~\cite{Joyce:2014kja}, and scalar waves might play an important role in such scenarios, even if their amplitude at emission has been shown numerically to be suppressed~\cite{Dar:2018dra}.

For an observer located far away from the source, in the radiation zone, the contributions of $A_\munu$ and $B$ to the field equations can be neglected as we adopted a Brans-Dicke field at large scales. 
This reduces the integrands to $S_\munu\rightarrow -16\pi \tau_\munu$ and $S\rightarrow -8\pi \tau_s$. The assumption of Brans-Dicke behaviour at leading order on cosmological scales can be motivated by the effective field theory (EFT) approach to dark energy~\cite{Creminelli:2008wc,Park:2010cw,Bloomfield:2011np,Gubitosi:2012hu,Bloomfield:2012ff,Gleyzes:2013ooa,Gleyzes:2014rba,Bellini:2014fua,Lombriser:2015cla,Frusciante:2019xia}. Under this formalism, the cosmological behaviour of Horndeski theories is described as an expansion in the metric perturbations and the scalar field represents the Goldstone boson of the symmetry breaking of time translation. The leading-order terms of a general Horndeski action at low energies are then equivalent to those of a Brans-Dicke theory with scalar field potential~\cite{Kennedy:2017sof,Kennedy:2018gtx,Kennedy:2019nie}.
Furthermore, we split the integration domain into two parts: the \emph{near zone} $\mathcal{N}$, which we assume to be screened, and the \emph{radiation zone} $\mathcal{C-N}$, assumed unscreened.
For the tensor wave, this splitting reads 
\begin{align}
\label{eq:MetricGreensIntergral}
h^{\mu\nu}(t,\bs{x}) = &\ 4 \int _\mathcal{N}\frac{\tau_{\mathcal{N}}^{\mu\nu}(t-|\bs{x}-\bs{x'}|, \bs{x}')}{|\bs{x}-\bs{x'}|} \dd^3\bs{x'} + 4\int _\mathcal{C-N}\frac{\tau_{\mathcal{C-N}}^{\mu\nu}(t-|\bs{x}-\bs{x'}|, \bs{x}')}{|\bs{x}-\bs{x'}|} \dd^3\bs{x'}\,.
\end{align}
Importantly, the effective stress-energy tensor takes different forms in the two different regions, represented by  $\tau_{\mathcal{N}}^{\mu\nu}$ and $\tau_{\mathcal{C-N}}^{\mu\nu}$ for the near and radiation zone, respectively. The latter is given by the Brans-Dicke effective stress-energy tensor in Eq.~\eqref{eq:GWSource}
and we describe $\tau_{\mathcal{N}}^{\mu\nu}$ in the following.

In the near zone, screening shall operate and we expect a close recovery of GR. However, while the effect of the scalar field is strongly attenuated, it nevertheless does not completely vanish. One can describe this contribution of the scalar field using the scaling method developed in Refs.~\cite{McManus:2016kxu,Lombriser:2016zfz,McManus:2017itv,Renevey:2020tvr}, in which the metric and the scalar field are expanded as 
\begin{align}
    \phi(t,\bs{x}) &= \phi_0 \left( 1 + \alpha^q \psi(t,\bs{x})+\mathcal{O}(\alpha^{2q})\right) \,,\\
    g_{\mu\nu}(t,\bs{x})&= g_{\mu\nu}^\0(t,\bs{x})+\alpha^q g_{\mu\nu}^\q(t,\bs{x})+\mathcal{O}(\alpha^{2q})\,.
\end{align}
Here, $\phi_0$ is the local background value of the scalar field sourced by the local environment.
It varies sufficiently slowly so as to be treated as a constant.
The dynamical contribution of the scalar field is captured by $\psi$.
The scaling parameter $\alpha$ is the relevant quantity describing the operation of a screening mechanism and was already introduced in Eq.~\eqref{eq:action_screened_horndeski}, where it represents the couplings to different terms in the scalar-tensor action.
Finally, the exponent $q\in\mathbb{R}$ is uniquely determined for each screening mechanism~\cite{McManus:2016kxu}.
In particular, $\alpha^q$ needs to be a small parameter, such that  $\alpha^q\psi\ll 1$, ensuring that the expansion remains viable.

The metric is decomposed into two parts. The first contribution $g_\munu^\0$ satisfies the Einstein field equations, where the local background scalar field effectively modifies the coupling to matter,
\begin{align}
   G_{\mu\nu}^\0 = \frac{8 \pi G}{\phi_0} T_{\mu\nu}\,. \label{eq:einstein_field_equation}
\end{align}
The second term $g_\munu^\q$ includes the leading contributions of the scalar field to the metric, encoded in the functional $A_\munu[g,\psi]$. Its equation of motion is found by substituting $\phi\rightarrow\phi_0(1+\alpha^q\psi)$ in Eq.~\eqref{eq:genericMetricFieldEquation} and keeping only the terms proportional to $\alpha^q$.
The equation for $\psi$ is found using the same substitution in Eq.~\eqref{eq:genericScalarFieldEquation}, where the functional $B[g,\psi]$ becomes the leading contribution together with the stress-energy tensor. In fact, $q$ is chosen such that the scalar field is sourced by the stress energy tensor and the field equations are convergent in the limit $\alpha^q\rightarrow 0$, which uniquely defines $q$.
The resulting expressions are
\begin{align}
G^\q_\munu=&\ \psi G_\munu^\0+\nabla_\mu\nabla_\nu\psi-g_\munu^\0\Box\psi+\delta_C\,\frac{1}{2}\,g_\munu^\0 V_0\phi_0^{n_V-1}\psi^{n_V}
\nonumber\\
&+\delta_K\,\phi_0^3\Gamma_2(\phi_0)\left(\frac{1}{2}\, g_\munu^\0 \tilde{X}^2+\tilde{X} \nabla_\mu\psi\nabla_\nu\psi\right)
    \nonumber\\
    &+ \delta_V\,\phi_0^2 \Gamma_3(\phi_0)\left(\frac{1}{2}\,g_\munu^\0\nabla^\alpha\psi\nabla_\beta\nabla_\alpha\psi\nabla^\beta\psi+\Box\psi\nabla_\mu\psi\nabla_\nu\psi- \nabla^\alpha\psi\nabla_\alpha\nabla_{(\mu}\psi\nabla_{\nu)}\psi\right)\, ,
\\[0.2cm]
    \frac{8\pi G}{\phi_0} \left(T - 2 \frac{\partial T}{\partial \phi}\right)=&\  \delta_C\, n_V\,\frac{\psi^{n_V-1}}{\phi_0^{n_V-1}}V_0+2\delta_K\,\phi_0^3\,\Gamma_2(\phi_0)\big(\tilde{X}\Box\psi+\nabla_\mu \tilde{X}\nabla^\mu \psi\big)\nonumber\\&+\delta_V\,\phi_0^2\,\Gamma_3(\phi_0)\big((\Box\psi)^2-\nabla_\mu\nabla_\nu\psi\nabla^\mu\nabla^\nu\psi\big)\,,\label{eq:effective_scalar_equation}
\end{align}
where $\tilde{X}=-1/2g^\0_\munu \partial^\mu\psi\,\partial^\nu\psi$ and the covariant derivatives as well as the contractions are obtained using $g_\munu^\0$. Allowing for an embedding of the chameleon, k-mouflage, and Vainshtein screening mechanisms, we have assumed that $V(\phi)=V_0\alpha^{n_V q}\psi^{n_V}$ with $0<n_V<1$, $\Gamma_2(\phi)=\Gamma_2(\phi_0)\neq 0$ and $\Gamma_3(\phi)=\Gamma_3(\phi_0)\neq 0$ at leading order in $\alpha^q\psi$~\cite{McManus:2016kxu,Lombriser:2016zfz,McManus:2017itv,Renevey:2020tvr,Lombriser:2014dua}.
For the remainder of this work, we shall furthermore adopt a normalisation of the gravitational constant by $G/\phi_0=1$. By looking at the right-hand side of Eq.~\eqref{eq:effective_scalar_equation}, one can see that in the case of k-mouflage and Vainshtein screening, the scalar contribution enters as a total derivative. In these cases, this holds true for any matter distribution and therefore only the total mass of the objects can affect the scalar field. Under the scaling formalism, one can expand the near-zone effective stress-energy tensor $\tau^\munu_\mathcal{N}$ as 
\begin{align}
    \tau^\munu_\mathcal{N}=\tau^\munu_\mathrm{GR}+\alpha^q\tau^\munu_\q+\mathcal{O}(\alpha^{2q})\,,
\end{align}
where $\tau^\munu_\mathrm{GR}$ is the part fully described by GR and $\tau^\munu_\q$ will be calculated to first Post-Newtonian order (1PN) in Sec.~\ref{sec:Correction_Metric_Near_Zone}. We also note that in both regions, $\mathcal{N}$ and $\mathcal{C-N}$, the effective stress-energy tensor is conserved to leading order in $\alpha$, i.e., $\tau^{\mu\nu}{}_{,\mu}=0+\mathcal{O}(\alpha)$.

The gravitational waveform $h^{ij}$ can be decomposed into three parts,
\begin{align}
h^{ij}(t,\bs{x}) = h^{ij}_{(0)}(t,\bs{x})+ h^{ij}\e{BD}(t,\bs{x})+ \alpha^q h_{(q)}^{ij}(t,\bs{x}) + \mathcal{O}(\alpha^{2q}) \,.
\end{align}
Hereby, $h^{ij}_\0$ denotes the contribution that is described by GR, corresponding to the result of the retarded integral of $\tau^\munu_\textrm{GR}$ in the near zone and the GR part of $\tau_\mathcal{C-N}^\munu$ in the radiation zone.
This term has been computed to 2PN beyond the quadrupole formula in WW96.
In Sec.~\ref{sec:waveform_fully_screened}, we also calculate the contribution from the scalar part of $\tau_\mathcal{C-N}^\munu$ to 2PN, namely $h^{ij}_\textrm{BD}$. The last term, $\alpha^q h_{(q)}^{ij}$, originates from the screened effect of the near-zone scalar field and it will be derived to 1PN in Sec.~\ref{sec:corrections_to_waveform_from_scalar_field}.
In models where gravitational objects can be self-screened relative to an ambient scalar field, such as is the case for the chameleon mechanism, black holes are in fact insensitive to this extra degree of freedom \cite{Zhang:2017srh,Liu:2018sia}. Such behaviour motivates the distinction between the fully screened case where the near-zone contribution $\alpha^q h^{ij}_\q$ is neglected and the case where it is not.

\section{Gravitational waveform of the fully screened theory}
\label{sec:waveform_fully_screened}

We shall first consider the case where a system is completely screened in the near zone, i.e., it is fully described by GR in this region, and reduces to Brans-Dicke theory in the unscreened radiation zone.
We determine the gravitational waveform up to 2PN for an arbitrary source distribution in this scenario. Hereby, we differentiate between the waveform described by GR and the extra terms from the Brans-Dicke scalar field, namely 
$h^{ij}=h^{ij}_{(0)}+h^{ij}\e{BD}$ with $\alpha^q h_{(q)}^{ij}\rightarrow 0$. The solution $h_{(0)}^{ij}$
can be found in WW96 and we calculate $h^{ij}\e{BD}$. In Sec.~\ref{sec:FullyScreenedMetricInTheNearZone}, we find the instantaneous metric and stress-energy tensor in the near zone. In Sec.~\ref{sec:FullyScreenedMetricInTheRadiationZone}, we derive the metric and scalar field in the radiation zone as sourced by the near zone. Finally in Sec.~\ref{sec:FullyScreenedFindingTheWaveform}, we calculate the waveform contribution from the Brans-Dicke scalar field.

\subsection{Metric and effective stress-energy tensor in the near zone}
\label{sec:FullyScreenedMetricInTheNearZone}

To compute the gravitational waveform to 2PN beyond the quadrupole approximation, one must determine the gravitational field in the near zone.
The contributions made to the waveform by the metric in the near zone are twofold.
First, they directly act as a source for higher-order contributions to the gravitational stress-energy tensor as $\Lambda^{\mu\nu} \in \tau^{\mu\nu}$. See the first term in Eq.~\eqref{eq:MetricGreensIntergral}.
Secondly, they contribute indirectly by sourcing the metric at a generic point in the radiation zone, which in turn is used in the second part of  Eq.~\eqref{eq:MetricGreensIntergral}.
This can be understood as the light cone being modified by the gravitational field, changing the propagation of the gravitational wave.

For a many-body system, we define its characteristic size $\mathcal{S}=\max\{ r_{AB}=|\bs{x}_A-\bs{x}_B|, \forall A,B \}$ as being the largest distance between any pair of point particles. The radius $\mathcal{R}$ of the near zone is defined to match the scale below which the screening mechanism is effective and where we assume that GR applies exactly. We will also refer to it as the screening radius. Since any object in the system lies well within the screening radius, $\mathcal{R}\gg\mathcal{S}$, the retardation of the fields is negligible at short scale and time derivatives can be treated as a higher-order term as in the standard PN approximation. In other words, the metric can be found in terms of instantaneous potentials.
We denote the constant spacelike hyper-surface at time $u=t-R$ as $\mathcal{M}$, with $R$ the distance between the source and the observer, and it is bounded by a sphere of radius $\mathcal{R}$.
We have assumed that the near zone is fully screened and, hence, that the metric field equations in the screened region are equivalent to those of GR (see Sec.~\ref{sec:modified_equations_with_screening}).
We skip the calculation of the metric and reproduce here the results of Sec.~III of WW96. The gravitational field components are given by
\begin{subequations}
\label{eq:nearZoneh}
\begin{align}
h^{00} =&\, 4\left(  U + \frac{1}{2}\partial_t^2 \mathcal{X} - P + 2 U^2 \right)+ \mathcal{O}(\epsilon^{5/2}) \,,\\
h^{0i} =&\, 4  U_i + \mathcal{O}(\epsilon^{5/2})\,,\\
h^{ij} =&\, 4 P_{ij} + \mathcal{O}(\epsilon^{5/2}) \,,
\end{align}
\end{subequations}
where we recall that
$\epsilon = v^2/c^2$ and for the background metric
\begin{subequations}
\label{eq:nearZoneg}
\begin{align}
g^{00} &= -(1 + 2  U + \partial_t^2 \mathcal{X} + 2  U^2) + \mathcal{O}(\epsilon^{5/2})\,, \\
g^{0i} &= -4  U_i + \mathcal{O}(\epsilon^{5/2}) \,, \\
g^{ij} &= (1 - 2 U -  \partial_t^2 \mathcal{X})\delta^{ij} + \mathcal{O}(\epsilon^{2}) \,,
\end{align}
\end{subequations}
from which the determinant is found to be 
\begin{equation}
(-g) =1+4(U+\frac{1}{2}\partial_t^2 \mathcal{X}) - 8(P-U^2) + \mathcal{O}(\epsilon^{5/2})\,.
\end{equation}
The instantaneous potentials are defined as
\begin{subequations}
\begin{align}
\label{eq:instPotential1}
U(u,\bs{x}) &\equiv \int_\mathcal{M}\frac{\dd^3\bs{x'}}{|\bs{x} - \bs{x}'|} (T^{00} + T^{ii}) (u, \bs{x'})
\,,\\
\label{eq:instPotential2}
\mathcal{X}(u,\bs{x}) &\equiv \int_\mathcal{M}\dd^3\bs{x'}|\bs{x} - \bs{x}'| (T^{00} + T^{ii}) (u, \bs{x'})
\,,\\
\label{eq:instPotential3}
U_{i}(u,\bs{x}) &\equiv \int_\mathcal{M}\frac{\dd^3\bs{x'}}{|\bs{x} - \bs{x'}|}  T^{0i} (u, \bs{x'})
\,,\\
\label{eq:instPotential4}
P_{ij}(u,\bs{x}) &\equiv \int_\mathcal{M}\frac{\dd^3\bs{x'}}{|\bs{x} - \bs{x'}|} \left[ T^{ij} + \frac{1}{4\pi}\left(U_{,i}U_{,j}-\frac{1}{2}\delta_{ij} U_{,k}U_{,k}\right)\right](u, \bs{x'})
\,.
\end{align}
\end{subequations}

With the metric in the near zone found to $\mathcal{O}(\epsilon^{5/2})$, the stress-energy tensor can also be found to $\mathcal{O}(\epsilon^{5/2})$, which will be required to determine the metric in the radiation zone.
The stress-energy tensor is given by
\begin{subequations}
\label{eq:NearZoneTExpansion}
\begin{align}
T^{00} =& \sum_A m_A \big[ 1 -  U + \frac{1}{2}\partial_t^2 \mathcal{X} + \frac{1}{2}v_A^2 + \frac{1}{2}U^2 + \frac{3}{2}U v_A^2 \nonumber \\
&+ 4 P + \frac{3}{8} v_A^4 - 4 U_i v^i_A + \mathcal{O}(\epsilon^{5/2}) \big]\delta^3(\bs{x}- \bs{x}_A(t))  \,,\\
T^{ij} =& \sum_A m_A v_A^i v_A^j \big[ 1 -  U + \frac{1}{2}v_A^2 + \mathcal{O}(\epsilon^2) \big]\delta^3(\bs{x}- \bs{x}_A(t))  \,,\\
T^{0j} =& \sum_A m_A v_A^j \big[ 1 - U - \frac{1}{2}\partial_t^2 \mathcal{X} + \frac{1}{2}v_A^2 + \mathcal{O}(\epsilon^{5/2}) \big]\delta^3(\bs{x}- \bs{x}_A(t)) \,.
\end{align}
\end{subequations}
We will only need the source $\tau^{\mu\nu}$ for the calculations performed here
to the orders $\mathcal{O}(\rho\epsilon)$ for $\tau^{00}$, $\mathcal{O}(\rho\epsilon)$ for $\tau^{ij}$, and $\mathcal{O}(\rho\epsilon^{3/2})$ for $\tau^{0i}$, where $\rho\sim\mathcal{O}(\epsilon)$.
With the background stress-energy tensor determined by Eq.~\eqref{eq:NearZoneTExpansion}, we are left with finding $\Lambda^{\mu\nu}$ to the appropriate orders. The non-compact part of the effective stress-energy tensor in Eq.~\eqref{eq:GravitationalStressEnergy} with the near-zone metric of Eqs.~\eqref{eq:nearZoneh} and \eqref{eq:nearZoneg} at the background level is given by
\begin{subequations}
\label{eq:NearZoneLambdaExpansion}
\begin{align}
\Lambda^{00} =&\,  14 (U + \frac{1}{2} \partial_t^2 \mathcal{X})_{,k}^2 + 16 [-U\ddot{U} + U_{,k}\dot{U}_{k} - 2 U_{k}\dot{U}_{,k} + \frac{5}{8} \dot{U}^2 \nonumber \\
&+ \frac{1}{2} U_{m,k}(U_{m,k} + 3 U_{k,m} ) + 2 P_{,k}U_{,k} - P_{kl} U_{,kl} \nonumber \\
&- \frac{7}{2} U U_{,k}^2 ]
+ \mathcal{O}\left(\rho\epsilon^{3}\right)\,,\\
\Lambda ^{0 i} =&\, 16 [(U+\partial_t^2 \mathcal{X})_{,k}(U_{k,i} - U_{i,k}) + \frac{3}{4}\dot{U}(U+\partial_t^2 \mathcal{X})_{,i}] 
+ \mathcal{O}\left(\rho\epsilon^{5/2}\right)\,,\\
\Lambda ^{ij} =&\, 4[(U+\partial_t^2 \mathcal{X})_{,i}(U+\partial_t^2 \mathcal{X})_{,j} - \frac{1}{2}\delta_{ij}(U+\partial_t^2 \mathcal{X})_{,k}(U+\partial_t^2 \mathcal{X})_{,k}] \nonumber \\
&+16[ 2U_{,(i}\dot{U}_{j)} - U_{k,i}U_{k,j} - U_{i,k}U_{j,k} + 2 - U_{k,(i}U_{j),k} \nonumber \\
&- \delta_{ij}(3/2 \dot{U}^2 + U_{,k}\dot{U}_k - U_{m,k}U_{[m,k]}]
+ \mathcal{O} \left(\rho\epsilon^{5/2}\right)\,.
\end{align}
\end{subequations}
We do not need $\Lambda_s$ in this region as we have assumed that it vanishes due to screening.
In Sec.~\ref{sec:FullyScreenedMetricInTheRadiationZone}, we will also need $\tau_s$ evaluated in the near zone. Neglecting the effect of the near-zone scalar field, i.e., $\varphi\rightarrow 1$, the scalar source term substantially simplifies to
\begin{align}
\tau_s =& - \frac{\sqrt{-g}}{(3+2\omega_0)}\left(T - 2  \frac{\partial T}{\partial \phi}\right) \nonumber \,\\
=& \frac{1}{(3+2\omega_0)} \sum_A m_A\left( 1-2\delta_Cs_A - \frac{1}{2} v_A^2 + U\right)\delta^3(\bs{x}- \bs{x}_A(t))\,.
\end{align}

\subsection{Metric in the radiation zone}
\label{sec:FullyScreenedMetricInTheRadiationZone}

At leading order, there is no contribution to the gravitational waveform from the radiation zone.
This is due to the leading order GW being sourced by the compact matter component, which lies solely in the near zone.
However, as we calculate the gravitational wave to higher orders, the non-compact stress-energy components, $\Lambda^{ij}$ and $\Lambda^{ij}_s$ in Eq.~\eqref{eq:GWSource}, contribute to the waveform through the scattering of gravitational waves on the background. 
As such, we need to compute the metric in the radiation zone as sourced by the near zone, which we will denote by $h_\mathcal{N}^{\mu\nu}$ in line with Refs.~\cite{Mirshekari:2013vb,Lang:2013fna}. To find $h_\mathcal{N}^{\mu\nu}$,
we compute the retarded integral 
\begin{equation}
\label{eq:RadZoneNearIntergral}
h_\mathcal{N}^{\mu\nu}(t', \bs{x}') = 4 \int _\mathcal{N'}\frac{\tau^{\mu\nu}(t'-|\bs{x}'-\bs{x}''|, \bs{x}'')}{|\bs{x}'-\bs{x}''|} \dd^3\bs{x}''\,.
\end{equation}
Since the integration is performed over the screened region, one should expect that the results that we thereby obtain mimic those of GR. Note that the domain $\mathcal{N'}$ of the integral is slightly different from the near zone $\mathcal{N}$ as it is part of the past light cone of another field point, $x^{\prime\mu}$ instead of $x^\mu$. The main difference resides in the modification of the time component in the near zone, which becomes $x^{\prime 0}=t'-|\bs{x}'-\bs{x}''|$ and $\bs{x}''\in\mathcal{N}'$. We define the distance and the direction of the arbitrary field point $x^{\prime\mu}$ as $R'$ and $\hat{N}'^{i} = x'{}^i / R'$, respectively.
The metric at the detector is then found by taking $R' = R$ and the detector lies in the direction $\hat N$ from the source.

In the screened region, the source $\tau^{\mu\nu}$ takes the same form as in GR and hence the metric solution in the radiation zone is the same as in WW96. As we will see, deviations from GR still arise from the radiation zone. We omit the details of such a calculation and restrict the discussion here to an outline of how this computation is performed. The retarded integral~\eqref{eq:RadZoneNearIntergral} is evaluated over the domain $\mathcal{N}'$ and so it can be expressed as a sum of moments. 
This expansion is valid as long as the distance to the field point, $R'$, is much greater than the characteristic size of the system, $\mathcal{S}$, leading to an expansion in $R'{}^{-1}$. As such,
\begin{equation}
\label{eq:radiationMetricFromNearMoments}
h^{\mu\nu}(t,\bs{x}) = 4 \sum_{q=0}^\infty \frac{(-1)^q}{q!}\left(\frac{1}{R'}M^{\mu\nu k_1 \ldots k_q}\right)_{,k_1 \ldots k_q}\,, 
\end{equation}
where
\begin{equation}
M^{\mu\nu k_1 \ldots k_q} = \int_\mathcal{M'}\tau^{\mu\nu} x'^{k_1}\ldots x'^{k_q}\dd^3x' \,.
\end{equation}
We evaluate these moments across a constant time hyper-surface $\mathcal{M'}$ at the retarded time $u = t' - R'$. Since the integral is still performed over the near zone, the effective stress-energy tensor takes its GR form and $\Lambda_s$ does not contribute.

To 2PN beyond the quadrupole formula only the compact part of $\tau^{\mu\nu}$ contributes, where higher-order compact contributions arise through inserting the instantaneous potentials into the determinant $-g$ in Eq.~\eqref{eq:GWSource}.
This renders these integrals relatively simple because of the Dirac $\delta$-functions in the stress-energy tensor. Furthermore, many of the higher-order moments can be written in terms of the lower moments together with the momentum currents 
\begin{equation}
\mathcal{J}^{iQ} = \epsilon^{iab}\int_\mathcal{M}\tau^{0b}x^{aQ}\dd^3x\,,
\end{equation}
where $\epsilon^{iab}$ is the Levi-Civita symbol. Following the derivations in WW96 (Eq.~(5.5)) and Lang14 (Eq.~(4.33)), we find
\begin{subequations}
\begin{align}
h_\mathcal{N}^{00} =&\, 4\frac{\tilde{m}}{R'} + 7\left(\frac{m}{R'}\right)^2 +  2\left(\frac{M^{ij}}{R'}\right)_{,ij} - \frac{2}{3} \left(\frac{M^{ijk}}{R'}\right)_{,ijk} \,, \\
h_\mathcal{N}^{0i} =& - 2 \left(\frac{\dot{M}^{ij} - \epsilon^{ija}\mathcal{J}^a}{R'}\right)_{,j} + \frac{2}{3}\left( \frac{\dot{M}^{ijk} - 2\epsilon^{ika}\mathcal{J}^{aj}}{R'} \right)_{,jk} \,, \\
h_\mathcal{N}^{ij} =& \left(\frac{m}{R'}\right)^2 \hat{N}'^{ij} + 2 \frac{\ddot{M}^{ij}}{R'} - \frac{2}{3}\left( \frac{\ddot{M}^{ijk} - 4\epsilon^{(i|ka}\dot{\mathcal{J}}^{a|j)}}{R'} \right)_{,k} \,,
\end{align}
\end{subequations}
where $\hat{N}'^{ij}=\hat{N}'^i\hat{N}'^j$, all moments are evaluated on the constant $u'-R'$ hyper-surface $\mathcal{M'}$, $\tilde{m} = \sum_A m_A + \frac{1}{2}m_Av_A^2 - \frac{1}{2}m_A U_A$, and $U_A$ is the total gravitational potential at the point $\bs{x}_A$ neglecting the infinite self-energy contribution of body $A$.

We are left with finding the scalar field in the radiation zone, $\varphi_\mathcal{N}$, to 2PN.
The integral~\eqref{eq:PhiRetIntergral} evaluated over the near zone can also be expanded into the sum of moments,
\begin{equation}
\label{eq:ScalarMomentSum}
\varphi_\mathcal{N}(t,\bs{x}) = 2 \sum_{q=0}^\infty \frac{(-1)^q}{q!}\left(\frac{1}{R}M_s^{ k_1 \ldots k_q}\right)_{,k_1 \ldots k_q} 
\end{equation}
with
\begin{subequations}
\begin{equation}
M_s^{ k_1 \ldots k_q} = \int_\mathcal{M}\tau_s x'^{k_1}\ldots x'^{k_q}\dd^3x'\,,
\end{equation}
where $\tau_s$ is given by Eq.~\eqref{eq:tauScalar}.
The calculation is otherwise equivalent to that of the metric expansion. The $\mathcal{O}(\rho\epsilon^{3/2})$ moments are found to be
\begin{align}
M_s =&\, \frac{1}{(3+2\omega)}\sum_A m_A(1-2\delta_Cs_A - \frac{1}{2}v_A^2 +  U_A) \,,\\
M^i_s =&\, \frac{1}{(3+2\omega)}\sum_A m_A x_A^i (1-2\delta_Cs_A - \frac{1}{2}v_A^2 + U_A) \,,\\
M^{ij}_s =&\, \frac{1}{(3+2\omega)}\sum_A m_A x_A^{ij}(1-2\delta_Cs_A) \,,\\
M^{ijk}_s =&\, \frac{1}{(3+2\omega)}\sum_A m_A x_A^{ijk}(1-2\delta_C s_A) \,.
\end{align}
\end{subequations}
We will not need moments larger than the octopole.
Note that these differ from the results of Lang14 (Eq.~(6.11)) since the scalar part of the stress-energy tensor $\Lambda_s^{\mu\nu}$ 
is modified due to the screening effect. Because of this modification, when evaluated in the center of mass frame corrected to 2PN, the scalar dipole moment vanishes, unlike in the Brans-Dicke case.
We will see in Sec.~\ref{sec:FullyScreenedWaveformContributionFromTheRadiationZone} that this has a major effect on the waveform as many terms present in Brans-Dicke theory no longer contribute. The scalar field in the radiation zone is then determined by inserting these moments into Eq.~\eqref{eq:ScalarMomentSum}, 
\begin{equation}
\label{eq:phiRadZone}
\varphi_\mathcal{N}(t,\bs{x}') = 2\frac{M_s}{R'} - 2\left(\frac{M_s^{i}}{R'}\right)_{,i} + \left(\frac{M_s^{ij}}{R'}\right)_{,ij} - \frac{1}{3} \left(\frac{M_s^{ijk}}{R'}\right)_{,ijk} \,.
\end{equation}

\subsection{Finding the waveform}
\label{sec:FullyScreenedFindingTheWaveform}

With the metric 
in the near and radiation zones at hand, we are fully equipped to compute the gravitational waveform. As we are only interested in the metric waves, we focus our attention on the retarded integral of the gravitational field in Eq.~\eqref{eq:hRetIntergral} to 2PN beyond the quadrupole approximation. Moreover, the calculation applies to a field point where $R\gg\mathcal{R}\gg\mathcal{S}$. 
As a result, we may discard terms that decay faster than $1/R$  and will refer to this region of spacetime as the far zone to distinguish it from the radiation zone.

In Sec.~\ref{sec:FullyScreenedWaveformContributionFromTheNearZone}, we outline how the near zone contributes to the gravitational wave.
No further calculation is needed as the source in the near zone is equivalent to that of GR and we can use the results of WW96.
In Sec.~\ref{sec:FullyScreenedWaveformContributionFromTheRadiationZone}, we discuss the contribution from the radiation zone. 
As the metric found from $\Lambda^{ij}$ again takes a form equivalent to GR, we refer to WW96 for the bulk of the calculation.
However, unlike in GR, the wave is also sourced by $\Lambda_s^{ij}$, which leads to new terms.

\subsubsection{Waveform contribution from the near zone}
\label{sec:FullyScreenedWaveformContributionFromTheNearZone}

The calculation of the gravitational waveform to 2PN beyond the quadrupole at the detector is a continuation of the calculation performed in Sec.~\ref{sec:FullyScreenedMetricInTheRadiationZone}. 
The main difference is that now we only need to keep terms in the metric expansion that decay as $R^{-1}$ since terms decreasing faster are negligible if the observer lies sufficiently far from the source. Furthermore, as we are evaluating the waveform and not the whole metric, we must remember that only the spatial part of the metric solution is needed and perform a transverse and traceless projection.
This projection moreover cancels terms proportional to either $\delta_{ij}$ or the directional vector to the detector $\bs{N}$.
The calculation of the gravitational waveform due to the near zone is analogous to WW96, and so we refer to Ref.~\cite{Will:1996zj} for the details, only providing here a brief overview of the derivation.

The moments in the expansion~\eqref{eq:radiationMetricFromNearMoments} can be expanded in $R^{-1}$, and 
since we are calculating the metric in the far zone, only the leading order is required.
The expansion can then be expressed in terms of \emph{Epstein-Wagoner} moments \cite{Epstein:1975pos}
by employing the conservation of the effective stress-energy tensor.
This simplifies Eq.~\eqref{eq:radiationMetricFromNearMoments} to 
\begin{equation}
\label{eq:EWExpansion}
h^{ij}_\mathcal{N} = \frac{2}{R} \frac{\dd^2}{\dd t^2}\sum^\infty_{m=0} \hat{N}_{k_i}\ldots\hat{N}_{k_m} I_\emph{EW}^{i j k_1 \ldots k_m} \,,
\end{equation}
where 
\begin{subequations}
\begin{align}
I^{ij}_\emph{EW} =& \int_\mathcal{M} \tau^{00}x^i x^j \dd^3 \bs{x} + I^{ij}_\emph{EW(surf)} \,, \\
I^{ijk}_\emph{EW} =& \int_\mathcal{M} (2 \tau^{0(i}x^{j)} x^k - \tau^{0k} x^i x^j ) \dd^3 \bs{x} + I^{ijk}_\emph{EW(surf)} \,, \\
I^{ij k_1 \ldots k_m}_\emph{EW} =&\, \frac{2}{m!} \frac{\dd^{m-2}}{\dd t^{m-2}} \int_\mathcal{M} \tau^{ij}x^{k_1} \ldots x^{k_m} \dd^3 \bs{x} \,, \quad (m\geq2)\,,
\end{align}
\end{subequations}
and
\begin{subequations}
\begin{align}
\frac{\dd^2}{\dd t^2}I^{ij}_\emph{EW(surf)}  =& \oint_{\partial\mathcal{M}} (4 \tau^{l(i} x^{j)} - (\tau^{kl}x^i x^j)_{,k}) \mathcal{R}^2 \hat{n}^l \dd^2\Omega\,, \\
\frac{\dd}{\dd t}I^{ijk}_\emph{EW(surf)} =& \oint_{\partial\mathcal{M}} (2 \tau^{l(i}x^{j)} x^k  - \tau^{kl}x^i x^j) \mathcal{R}^2 \hat{n}^l \dd^2\Omega
\end{align}
\end{subequations}
with $\hat{n}$ denoting a radially outward-directed unit vector. 

The explicit calculation of these moments is
an involved task,
making up the bulk of both the WW96 and Lang14 articles.
Due to the similarity
between the calculation performed in WW96 and here, we will not replicate it. In brief, $\tau^{ij}$ is expressed in terms of the instantaneous potentials using the results in Eqs.~\eqref{eq:NearZoneTExpansion} and \eqref{eq:NearZoneLambdaExpansion}. The potentials are themselves found using the matter distribution in Eq.~\eqref{eq:MatterDistribution}.
Finally, the Epstein-Wagoner moments are inserted in the expansion \eqref{eq:EWExpansion}.

The solutions for the Epstein-Wagoner moments are given in Eqs.~6.6 of WW96.
However, in their calculation of the gravitational waveform, terms from the near-zone and the radiation-zone contributions that depend on $\mathcal{R}$ cancel out. 
There is no guarantee for this in our model and so we highlight the terms proportional to $\mathcal{R}$,
\begin{subequations}
\label{eq:EWBoundaryContribution}
\begin{align}
\label{eq:EWBoundaryContributionA}
I^{ij}_\emph{EW} \supset & - \frac{14}{5} m \mathcal{R} \ddot{M}^{ij} \nonumber \\
=& - \frac{28}{5} m \mu \mathcal{R} \big( v^{ij} - x^{(i}a^{j)}\big)\,,\\
\label{eq:EWBoundaryContributionB}
I^{ijkl}_\emph{EW} \supset & - \frac{8}{35} m \mathcal{R} \ddot{M}^{ij} \delta^{kl} \nonumber \\
=& - \frac{8}{35} m \mu \mathcal{R}  \big( v^{ij} - x^{(i}a^{j)}\big) \delta^{kl}\,,\\
\label{eq:EWBoundaryContributionC}
I^{ijklmn}_\emph{EW} \supset & - \frac{2}{315} m \mathcal{R} \ddot{M}^{ij} \delta^{(kl} \delta^{mn)} \nonumber \\
=& - \frac{2}{315} m \mu \mathcal{R} \big( v^{ij} - x^{(i}a^{j)}\big) \delta^{(kl} \delta^{mn)}\,.
\end{align}
\end{subequations}
These terms depend on the size of the near zone $\mathcal{R}$, which we have set to the screening radius. 
In GR these terms were expected not to appear in the final waveform as the value of $\mathcal{R}$ is unphysical.
Their cancellations with the boundary terms of the radiation-zone integration is checked in the following.

\subsubsection{Waveform contribution from the radiation zone}
\label{sec:FullyScreenedWaveformContributionFromTheRadiationZone}

When determining the metric in the radiation zone in Sec.~\ref{sec:FullyScreenedMetricInTheRadiationZone}, we found that the metric sourced by the near zone remains equivalent to that in GR.
Deviations in the metric occur due to the self-interaction in the radiation zone.
The modification of the gravitational theory in the radiation zone by Brans-Dicke gravity
only begins to affect the metric of this region at $\mathcal{O}(\epsilon^2)$. 

We first want to know if the extra terms \eqref{eq:EWBoundaryContributionA}-\eqref{eq:EWBoundaryContributionC} remain in the gravitational waveform or are cancelled by contributions from the radiation zone. The relevant part of $\Lambda^{ij}$ in the radiation zone up to $\mathcal{O}(\epsilon^{3})$ is found to be
\begin{equation}
\label{eq:metricTailTerm}
\Lambda^{ij} \supset - h^{00}_{(1)}\ddot{h}^{ij}_{(2)} + \frac{1}{4}h^{00}_{(1)}{}_{,(i}h^{00}_{(2)}{}_{,j)} + \frac{1}{2}h^{00}_{(1)}{}_{,(i}h^{kk}_{(2)}{}_{,j)} + 2h^{00,(i}_{(1)} \dot{h}^{j)0}_{(2)}\,,
\end{equation}
where the subscripts denote the PN orders used.
%
We may use the results in Eq.~(5.8) of WW96 for the metric components $h^{ij}$ to 2PN beyond the quadrupole order, 
\begin{equation}
h^{ij} \supset \frac{4m}{R}\int_0^\infty \dd s\, \partial_t^4M^{ij}(u-s)\left[\ln\left(\frac{s}{2R+s}\right) + \frac{11}{12}\right]\,.
\end{equation}
This is the first tail term arising from the interaction of the gravitational wave with the background Newtonian potential and is of order 1.5PN. Therefore, only one more iteration is required to reach the desired precision of 2PN.
The components of the gravitational stress-energy tensor that we need to consider are
\begin{equation}
\Lambda^{ij} \supset - h^{00}_{(1)}\ddot{h}_{(1.5)}^{ij} + \frac{1}{2}h^{00}_{(1)}{}_{,i} h^{00}_{(1.5)}{}_{,j} + \frac{1}{2}h^{00}_{(1)}{}_{,(i}h^{kk}_{(1.5)}{}_{,j)} \,.
\end{equation}
As can be seen with Eq.~\eqref{eq:EWExpansion}, the first term, the source is proportional to $\delta_{ij}$ and so is not transverse-traceless.
For the second and third terms, the monopole-monopole couplings scale as $R^{-1}{}_{,(i}R^{-1}{}_{,j)}$, and when the derivatives are evaluated, they have an angular dependence of the form $\hat{N}^{ij}$ , which is not transverse-traceless. 
Hence, only the terms from monopole-quadrupole, and monopole-current quadrupole are relevant in $\Lambda^{ij}$.
The metric components due to the near zone again mimic those of GR, leading to the following result,
\begin{align}
h^{ij} \supset & \frac{4m}{3R}\hat{n}^k \int^\infty_0 \dd s\, \partial_t^5 M^{ijk}(u-s)\left[\ln\left(\frac{s}{2R+s}\right) + \frac{97}{60}\right] \nonumber \\
& - \frac{16 m}{3 R} \epsilon^{(i|ka}\hat{n}^k \int^\infty_0 \dd s\, \partial_t^4 J^{a|j)}(u-s)\left[\ln\left(\frac{s}{2R+s}\right) + \frac{7}{6}\right] \nonumber \\
& + \frac{1912}{315}\frac{m}{R}\partial_t^4 M^{ij}(u) \mathcal{R} \,,
\end{align}
which is the remainder of Eq.~(5.8) in WW96. We can check that the extra contributions linear in $\mathcal{R}$ from the boundary exactly cancel the contributions from Eq.~\eqref{eq:EWBoundaryContribution}.
This means that to 2PN, the radius $\mathcal{R}$ does not enter the waveform as in GR.

The remaining contributions arise from
$\Lambda_s^{ij}$ in Eq.~\eqref{eq:LambdaS}.
The calculation is similar to that performed in Sec.~VI.C of Lang14, but with differing prefactors.
In Sec.~\ref{sec:FullyScreenedMetricInTheRadiationZone}, we found that the spatial part of the gravitational field $h^{ij}$ scales as $R^{-2}$ to leading order, leaving no contribution at the quadrupole order to the gravitational waveform in the far zone. For the next order, we expand $\Lambda_s$ to 0.5PN beyond the quadrupole to find 
\begin{equation}
\tau^{ij}_\textrm{BD} = \frac{\pi}{16} (3+2\omega) \varphi\e{mono}^{,(i}\varphi\e{di}^{,j)}+\mathcal{O}(\epsilon^2)\,.
\end{equation}
We denote by $\varphi\e{mono}$, $\varphi\e{di}$, $\varphi\e{quad}$ and $\varphi\e{oct}$, the monopole, dipole, quadrupole and octopole contributions to $\varphi$ in Eq.~\eqref{eq:phiRadZone}, respectively. 
The subscript BD refers to the corrections from the Brans-Dicke field.
Since the dipole term vanishes in the centre of mass frame, $\Lambda_s$ does not contribute at this order. At 1PN, we find 
\begin{equation}
\tau^{ij}_\textrm{BD} = \frac{\pi}{16} (3+2\omega) \big( 
  \varphi\e{mono}^{,(i}\varphi\e{quad}^{,j)}
+ \varphi\e{mono}^{,(i}\varphi\e{mono}^{,j)}
\big)+\mathcal{O}(\epsilon^{5/2})\,,
\end{equation}
where 
the cross-term monopoles are evaluated at $\mathcal{O}(\epsilon)$. The expansion in terms of the moments of the scalar field is analogous to that in Eq.~(6.29) of Lang14
and leads to the 1.5PN contribution to the waveform
\begin{align}
\label{eq:ScalarContribution1}
P^{1.5}h^{ij}_\textrm{BD} = \frac{4m_s}{ R}\bigg(-\frac{1}{12}\partial_t^3 M^{ij}_s\bigg)\,,
\end{align}
where $P^k$ stands for the $k$-th PN order beyond the quadrupole formula and $m_s=\sum m_A(1-2\delta_C s_A)$. This term enters at 1.5PN beyond the quadrupole and is an extra contribution to the waveform arising from the scalar quadrupole.  Interestingly, in pure Brans-Dicke theory there is a new tail term that arises from the dipole-dipole interactions, which however does not appear in our screened theory as in the near zone we are sourced by the GR solution for the metric.

Finally, to find the contribution to the metric attributed to $\Lambda_s$ at 2PN beyond the quadrupole, we need only to consider the monopole-octopole terms as all other contributions at this order vanish when taking the transverse-traceless projection,
\begin{equation}
\tau^{ij}_\textrm{BD} \supset \frac{\pi}{16} (3+2\omega) 
  \varphi\e{mono}^{,(i}\varphi\e{oct}^{,j)}+\mathcal{O}(\epsilon^3)\,,
\end{equation}
where $\varphi\e{oct}$ is the octopole term. Calculating the relevant contributions, we are left with the final 2PN term of the waveform, 
\begin{align}
\label{eq:ScalarContribution2}
P^2h^{ij}_\textrm{BD} = \frac{4 m_s}{ R}\left(-\frac{1}{60}\partial_t^4 M_s^{ija}\hat{N}^a \right)\,,
\end{align}
which enters at 2PN beyond the quadrupole.
Again, we lose the additional tail terms found in Brans-Dicke theory due to the vanishing scalar dipole moment. Therefore, the waveform up to 2PN beyond the quadrupole formula corresponds to that of GR with the additional contributions at 1.5PN and 2PN from $\Lambda_s^{\mu\nu}$ of the form
\begin{align}
    h^{ij}_\textrm{BD} = \frac{4m_s}{ R}\bigg(-\frac{1}{12}\partial_t^3 M^{ij}_s\bigg)+\frac{4 m_s}{R}\left(-\frac{1}{60}\partial_t^4 M_s^{ija}\hat{N}^a \right)+\mathcal{O}(P^{2.5}h^{ij}_\textrm{BD})\,.
\end{align}

\section{Leading corrections to the waveform from the near-zone scalar field}
\label{sec:corrections_to_waveform_from_scalar_field}

We have now established that the effect of a scalar field on the gravitational waveform from a fully screened source enters at 1.5PN beyond the quadrupole formula and is due to its radiation-zone contribution. In the following, we shall compute the leading correction of the near-zone scalar field on the waveform of a screened source. In a Brans-Dicke model with no screening, the near-zone scalar field affects the waveform formula at 1PN~\cite{Lang:2013fna}. We expect the screened scalar field to also impact the system at 1PN, but this effect should be smaller than in the unscreened model, attenuated by the scaling parameter $\alpha^q$. 
Thus, $\alpha^q$ is expected to suppress the new near-zone contributions to the waveform formula. In this work, we shall neglect higher-order contributions $\mathcal{O}(\alpha^{2q})$, and we limit the post-Newtonian calculations to 1PN.
In Sec.~\ref{sec:Correction_Metric_Near_Zone}, we find the scalar contributions to the instantaneous metric and stress-energy tensors in the near zone. Then in Sec.~\ref{sec:metric_scalar_corrections_radiation_zone}, we show that the radiation-zone self-interactions do not affect the gravitational waveform and we find the scalar corrections to the Epstein-Wagoner moments.

\subsection{Corrections to the metric and effective stress-energy tensors in the near zone}\label{sec:Correction_Metric_Near_Zone}

Similarly to the fully screened case in Sec.~\ref{sec:waveform_fully_screened}, we first need to find the near-zone metric $g_\munu$ and gravitational field $h^{\mu\nu}$ up to 1PN. These tensor fields are described by GR with the added corrections, $\alpha^q g^{\munu}_\q$ and $\alpha^q h^{\mu\nu}_\q$, respectively. From there, we may calculate the leading $\alpha$-correction to the stress-energy tensors $\tau^\munu$ and $\tau_s$ at the required PN order.

Starting with the metric, the 1PN formalism in the screened regime was introduced in Ref.~\cite{McManus:2017itv} and later extended to Horndeski theories with luminal propagation of gravitational waves in Ref.~\cite{Renevey:2020tvr}. One can use these results to write the correction to the metric field up to 1PN,
\begin{subequations}
\label{eq:correction_metric_2PN}
\begin{align}
\alpha^q g^\q_{00} &= \alpha^q(\psi+\tilde{\Phi}_1-3\mathcal{A}_\psi-\mathcal{B}_\psi+6\tilde{\Phi}_2+\Phi_{Sc})+ \mathcal{O}(\epsilon^{3})\,, \\
\alpha^q g^\q_{0i} &= \alpha^q\gamma_{0i} + \mathcal{O}(\epsilon^{5/2})\,, \\
\alpha^q g^\q_{ij} &= (-\alpha^q\,\psi)\delta_{ij}+ \mathcal{O}(\epsilon^2) \,.
\end{align}
\end{subequations}
The post-Newtonian retarded potentials involving the scalar field are of the form
\begin{align}
\tilde{\Phi}_1(\bs{x})&= -\frac{1}{4\pi}\int \frac{\psi(\bs{x}') v(\bs{x}')^2}{|\bs{x}-\bs{x}'|^3}\dd^3\bs{x}'\,,
&
\mathcal{A}_\psi(\bs{x}) &= -\frac{1}{4\pi}\int \frac{\psi(\bs{x}') [\bs{v}(\bs{x}')\cdot(\bs{x}-\bs{x}')]^2 }{|\bs{x} - \bs{x}'|^5}\dd^3\bs{x}'\,,
\\[0.2cm]
\tilde{\Phi}_2(\bs{x}) &= \int \frac{\rho(\bs{x}') \psi(\bs{x}')}{|\bs{x}-\bs{x}'|}\dd^3\bs{x}'\,,
&
\mathcal{B}_\psi(\bs{x}) &= -\frac{1}{4\pi}\int \frac{\psi(\bs{x}')  [\bs{a}(\bs{x}')\cdot(\bs{x}-\bs{x}')] }{|\bs{x} - \bs{x}'|^3}\dd^3\bs{x}'\,,
\end{align}
\begin{align}
\grad^2 \gamma_{0i}&=-1/2\, \partial_i\dot{\psi}\,,
\\[0.1cm]
\bs{\nabla}^2\Phi_{Sc}&=-\delta_C\frac{V_0}{2\phi_0^{-n_V}}\psi^{n_V}+\delta_K\frac{\phi_0^4}{2}\Gamma_2(\phi_0)(\grad\psi\cdot\grad\psi)^2-\delta_V\phi_0^3\Gamma_3(\phi_0)\bs{\nabla}^2\psi(\grad\psi\cdot\grad\psi)\,,
\end{align}
where $\grad$ is the three-gradient, $\grad^2$ is the three-Laplacian and the $\alpha$-order of the scalar field $\psi\sim\psi^\q$ is kept implicit throughout this work.

Notice that the scalar field $\psi$ appears abundantly in the 1PN correction to the metric, but we have not yet described its behaviour in the near zone, which we shall resolve now. We recall the effective scalar field equation in the screened region as exposed in Eq.~\eqref{eq:effective_scalar_equation},
\begin{align}
    8\pi \left(T-2\pdv{T}{\phi}\right)=&\  \delta_C\, n_V\,\frac{\psi^{n_V-1}}{\phi_0^{n_V-1}}V_0+2\delta_K\,\phi_0^3\,\Gamma_2(\phi_0)\big(\tilde{X}\Box\psi+\nabla_\mu \tilde{X}\nabla^\mu \psi\big)\nonumber\\&+\delta_V\,\phi_0^2\,\Gamma_3(\phi_0)\big((\Box\psi)^2-\nabla_\mu\nabla_\nu\psi\nabla^\mu\nabla^\nu\psi\big)\,,\label{eq:effective_scalar_equation2}
\end{align}
where $\tilde{X}\equiv-g_\0^\munu \partial_\mu\psi\,\partial_\nu\psi/2$ and all contractions or covariant derivatives are made using the GR metric $g^\0_\munu$. At first sight, the PN order of the scalar field appears to be inconsistent between the three different terms on the right-hand side of Eq.~\eqref{eq:effective_scalar_equation2}. This prevents us from considering a superposition of screening mechanisms. Since the leading PN order of the trace of the stress-energy tensor and its derivative is $\mathcal{O}(\epsilon)$, the leading PN order of the scalar dynamical part $\psi$ depends on the screening mechanism at play.
As discussed in Refs.~\cite{Renevey:2020tvr,Bolis:2018kcq}, the scaling parameter $\alpha$ actually carries a PN order, so that the full scalar field term $\alpha^q\psi$ has the leading order $\mathcal{O}(\epsilon)$. The PN order of the scalar field $\mathcal{O}(\epsilon^k)$ is determined through Eq.~\eqref{eq:effective_scalar_equation2}, which in turn determines the PN order of $\alpha^q\sim\mathcal{O}(\epsilon^{(1-k)})$. From now on, we refer to the leading PN contribution of the scalar field as $\psi_k$, such that $\psi=\psi_k+\mathcal{O}(\epsilon^{k+1})$. The field equation describing the behaviour of $\psi_k$ then reads
\begin{align}
  8\pi \sums_Am_A(1-2\delta_C s_A)\delta^3(\bs{x}-\bs{x}_A(t))=&-\delta_C\,n_V\,\phi_0^{n_V-1}\psi_k^{n_V-1}\,V_0\nonumber
   \\
   &-\delta_K\,\phi_0^3\,\Gamma_2(\phi_0)\bigg((\grad\psi_k\cdot\grad\psi_k)\bs{\nabla}^2\psi_k
   \nonumber\\[-0.2cm]
   &\qquad\qquad\qquad\qquad+ \grad(\grad\psi_k\cdot\grad\psi_k)\cdot\grad \psi_k\bigg)
   \nonumber\\[0.2cm]
   &-\delta_V\,\phi_0^2\,\Gamma_3(\phi_0)\big((\bs{\nabla}^2\psi_k)^2-\nabla_i\nabla_j\psi_k\nabla^i\nabla^j\psi_k\big)\,.\label{eq:scalar_equation_leading_PN}
\end{align}
Since $n_V<1$, the field equation for theories exhibiting the chameleon mechanism is ill-defined. This is due to the fact that we considered the matter content to be described by massive point-like objects. In chameleon screening, compact objects with high energy densities are screened against the ambient scalar field. In the point-like approximation, we assume vacuum everywhere, while the density of the body follows a Dirac delta distribution, which is incompatible with the screening mechanism at play. Because GWs that are presently detected originate from very compact objects, one can simply assume that the scalar degree of freedom is fully screened by the very high internal density $\psi_k\propto\rho^{1-n_V}\sim 0$  and we recover the case developed in Sec.~\ref{sec:waveform_fully_screened}.

We now have all the ingredients to proceed with the calculation of the $\alpha$-correction to the gravitational field $h^{(q)\mu\nu}$ as well as the effective stress-energy tensors $\tau^{(q)\mu\nu}$ and $\tau_s^\q$ in the near zone. Starting with the former, we can use Eq.~\eqref{eq:gravPotential} to deduce its correction term. However, at 1PN only the $0i$-components are affected,
\begin{subequations}
\label{eq:nearZoneh_correction}
\begin{align}
h^{(q)00}=&\ h^{(q)ij} = 0+ \mathcal{O}(\alpha^{-q}\epsilon^{2}) \,,\\[0.2cm]
h^{(q)0i} =&\ -\gamma_{0i}^\q + \mathcal{O}(\alpha^{-q}\epsilon^{5/2})\,.
\end{align}
\end{subequations}
Note that we included $\alpha^{-q}$ in the neglected PN orders since the scale parameter carries a PN order itself. Once we take the full correction term $\alpha^q h^{(q) ij}$, we recover the usual PN orders. The effective stress-energy tensors $\tau^{(q)\munu}$ and $\tau_s^q$ are then found by inserting Eqs.~\eqref{eq:correction_metric_2PN} and \eqref{eq:nearZoneh_correction} into Eqs.~\eqref{eq:GravitationalStressEnergy} and \eqref{eq:tauScalar}. For the tensor $\Lambda^\munu$, no contribution from the scalar field is present at 1PN. Furthermore,
corrections to $\Lambda_s^{\mu\nu}$ are of $\mathcal{O}(\alpha^{2q})$ and we neglect them. The compact part of the effective stress-energy tensor in terms of the scalar field and gravitational potentials can be computed at 1PN using Eq.~\eqref{eq:MatterDistribution}, which gives
\begin{subequations}
\label{eq:corretion_stress_energy_tensor}
\begin{align}
    T^{(q)00}=&\ \sums_A m_A\delta^3(\bs{x}-\bs{x}_A(t))\frac{5}{2}\psi_k+\mathcal{O}(\alpha^{-q}\rho\epsilon^{2})\,,
    \\
    T^{(q)0i}=&\ T^{(q)ij}=0+\mathcal{O}(\alpha^{-q}\rho\epsilon^{3/2})\,.
\end{align}
\end{subequations}
We notice that only the compact part  of the stress-energy tensor is modified at 1PN, which allows for a straightforward integration.
The 1PN contribution of the near-zone scalar field to the effective stress-energy tensors can therefore be obtained using Eqs.~\eqref{eq:GWSource} and \eqref{eq:tauScalar}:
\begin{subequations}
\label{eq:correction_tau_munu}
\begin{align}
    \tau^{(q)00}=&-\sums_A m_A\delta^3(\bs{x}-\bs{x}_A(t))\frac{\psi_k}{2}+\mathcal{O}(\alpha^{-q}\rho\epsilon^2)\,,\\
    \tau^{(q)0i}=&\,0+\mathcal{O}(\alpha^{-q}\rho\epsilon^{3/2})\,,\\
    \tau^{(q)ij}=&\,0+\mathcal{O}(\alpha^{-q}\rho\epsilon^2)\,,
\end{align}
\end{subequations}
and
\begin{align}
    \tau^{(q)}_s=&\frac{3}{2(3+2\omega_0)}\sums_A m_A\left(1-2\delta_C s_A\right)\delta^3(\bs{x}-\bs{x}_A(t))\psi_k+\mathcal{O}(\alpha^{-q}\rho\epsilon^2)\,.\label{eq:correction_tau_s}
\end{align}

\subsection{Metric and scalar field corrections in the radiation zone and contribution to the waveform}\label{sec:metric_scalar_corrections_radiation_zone}

The metric in the radiation zone is found using a sum of moments as defined in Eq.~\eqref{eq:radiationMetricFromNearMoments}. The leading corrections from the near-zone scalar field can be calculated from the stress-energy tensor $\tau_\munu^\q$ and, at 1PN, only the 00-component of the gravitational field is affected. The Dirac $\delta$-function in the compact part of the stress-energy tensor ensures that we can easily integrate the moments and find
\begin{align}
    h_\mathcal{N}^{(q)00}(t,\bs{x}')=-\frac{2}{R'}\sums_A m_A\psi_k(\bs{x}_A(t))+\mathcal{O}(\alpha^{-q}\rho\epsilon^{2})\,.
\end{align}
We recall that $R'$ is the distance between an arbitrary field point $x^{\prime \mu}$ in the radiation zone and the source. When taking the transverse-traceless gauge, this correction term vanishes and therefore does not contribute to the gravitational waveform.

One follows a similar strategy to find the corrected scalar field in the radiation zone. Its moment expansion is defined in Eq.~\eqref{eq:ScalarMomentSum} and at 1PN, we find
\begin{align}
    \varphi_\mathcal{N}^\q=\frac{3}{2(3+2\omega_0)R'}\sums_A m_A(1-2\delta_C s_A)\psi_k(\bs{x}_A(t))+\mathcal{O}(\alpha^{-q}\rho\epsilon^{2})\,.
\end{align}
The corrected radiation-zone scalar field is important to take into account the scalar wave. However, this correction only affects the gravitational wave beyond 1PN and we do not need to compute its contribution to the gravitational waveform. 

As we are restricting ourselves to the leading corrections at 1PN beyond the quadrupole formula, the radiation-zone contributions can be neglected. Therefore, we focus our attention to the near-zone effects and the Epstein-Wagoner moments. Only the 00-component of the stress-energy tensor is corrected at this order and this means that the near-zone scalar field does not significantly affect the moments beyond the quadrupole. Furthermore, the contributing part of $\tau_\munu^\q$ is compact and easy to integrate over the near zone. The result is simply
\begin{align}
    I_\mathrm{EW}^{(q)ij}=-\frac{1}{2}\sums_A x_A^ix_A^j m_A\psi_k(\bs{x}_A(t))\,.
    \label{eq:correction_EW_moment}
\end{align}
This term is to be added to the usual GR result of the Epstein-Wagoner moments and additional terms in the waveform emerge from a modification of the equations of motion of the source as will become clear in Sec.~\ref{sec:reduced_case_binary_black_hole}.

\section{Black hole binary waveform}
\label{sec:reduced_case_binary_black_hole}

In Secs.~\ref{sec:waveform_fully_screened} and \ref{sec:corrections_to_waveform_from_scalar_field}, we have derived the deviation from the. GR waveform for the $N$-body problem. 
However, the targeted sources of the LIGO-Virgo collaboration are the late inspirals of binary objects, composed of black holes and/or neutron stars.
In the following, we shall restrict our discussion to the case of black hole-black hole binaries on quasi-circular orbits. The quasi-circular assumption is well motivated by the continuing loss of angular momentum circularising the orbits~\cite{BlanchetReview} for the late inspiral.

In Sec.~\ref{sec:fully_screened_binary}, we treat the case of the fully screened source, which is described by GR and give the corrections to 2PN beyond the quadrupole approximation to the waveform from a Brans-Dicke scalar field in the radiation zone. In Sec.~\ref{sec:two_body_pb_correction}, we find the effect of the near-zone scalar field on the equations of motion of the two-body problem. In Sec.~\ref{sec:waveform_1PN_correction}, we give the contributions up to 1PN beyond the quadrupole formula of the screened near-zone scalar field to the gravitational waveform. Finally, in Sec.~\ref{sec:example_galileon_cosmology}, we calculate the order of magnitude of the correction in the fully screened case as well as in the case with an effective screened scalar field for the cubic Galileon model~\cite{Nicolis:2008in}.

\subsection{Fully screened behaviour at 2PN}
\label{sec:fully_screened_binary}

Since the scalar field in the near zone is first assumed to be fully screened with no effect on the binary system, we can use the equations of the two-body system in GR to describe the behaviour of the source to 2PN (see Appendix~\ref{app:CalculatingTheGravitationalWaveform} or Ref.~\cite{Will:2018bme}). We work in the centre of mass coordinates that are valid to second PN order. We define the usual two body variables; the total mass $m = m_1+m_2$, mass difference $\delta m=m_1-m_2$, reduced mass $\mu = m_1m_2 / m$, symmetric mass ratio $\eta = \mu / m$, relative position $\bs{x} = \bs{x}_1 - \bs{x}_2$, relative distance $r = |\bs{x}|$, unit vector $\bs{\hat{n}} = \bs{x}/ r$, relative velocity $\bs{v} = \dot{\bs{x}}$ and direction to the observer $\bs{\hat{N}}$. The GR
equations of motion 
to second PN order are presented in Eq.~\eqref{eq:PNAcceliration}.

With these quantities in the centre of mass frame and the equations of motion for the two-body system, we are now equipped to find the new contributions to the GW in terms of the Newtonian parameters. We recall that the sensitivity of black holes is $1/2$ and hence, the scalar field in the chameleon case leaves no contribution to the gravitational waveform. Up to 2PN beyond the quadrupole approximation, the gravitational waveform becomes
\begin{align}
h^{ij} =&\ h^{ij}\e{GR}+h^{ij}\e{BD}+\mathcal{O}(\alpha^q)\,,
\\[0.2cm]
h^{ij}_\textrm{GR}=&\ \frac{\mu}{2R}\left(Q^{ij}_\textrm{GR}+P^{0.5}Q^{ij}_\textrm{GR}+P^{1}Q^{ij}_\textrm{GR}+P^{1.5}Q^{ij}_\textrm{GR}+P^{2}Q^{ij}_\textrm{GR}+...\right)\,,
\\[0.2cm]
h^{ij}_\textrm{BD}=&\ (1-\delta_C)\frac{\mu}{2R}\left(P^{1.5}Q^{ij}_\textrm{BD}+P^{2}Q^{ij}_\textrm{BD}+...\right)\,,\label{eq:hij_BD_2PN}
\end{align}
where
\begin{align} 
P^{1.5}Q^{ij}_\textrm{BD}=&\ \frac{8m^2}{3(3+2\omega_0)r^2} \left[2\hat{n}^{(i}v^{j)}-\dot{r}\hat{n}^{i}\hat{n}^{j} \right]\,, \label{eq:fully_screened_P15}
\\[0.2cm]
P^{2}Q^{ij}_\textrm{BD}=&\ 
\frac{2m\,\delta m}{15(3+2\omega_0)r^2}\left(
-12(\bs{\hat{N}}\cdot \bs{\hat{n}})v^iv^j-24(\bs{\hat{N}}\cdot \bs{v})\hat{n}^{(i}v^{j)}+\frac{16}{r}\dot{r}(\bs{\hat{N}}\cdot \bs{v})\hat{n}^{(i}x^{j)}+
\right.
\nonumber\\[0.1cm]
&\left.
+\frac{12m}{r^2}(\bs{\hat{N}}\cdot \bs{\hat{n}})\hat{n}^{(i}x^{j)}+\frac{16\dot{r}}{r}(\bs{\hat{N}}\cdot \bs{\hat{n}})v^{(i}x^{j)}-\frac{16}{r}(\bs{\hat{N}}\cdot \bs{v})v^{(i}x^{j)}-\frac{2m}{r^3}(\bs{\hat{N}}\cdot \bs{\hat{n}})x^{i}x^{j}-
\right.
\nonumber\\[0.1cm]
&\left.
-\frac{15\dot{r}^2}{r^2}(\bs{\hat{N}}\cdot \bs{\hat{n}})x^{i}x^{j}+\frac{3v^2}{r^2}(\bs{\hat{N}}\cdot \bs{\hat{n}})x^{i}x^{j}+\frac{6\dot{r}}{r^2}(\bs{\hat{N}}\cdot \bs{v})x^{i}x^{j}+\frac{6m}{r^2}(\bs{\hat{N}}\cdot \bs{x})\hat{n}^{i}\hat{n}^{j}+
\right.
\nonumber\\[0.1cm]
&\left.
+\frac{16\dot{r}}{r}(\bs{\hat{N}}\cdot \bs{x})\hat{n}^{(i}v^{j)}-\frac{8}{r}(\bs{\hat{N}}\cdot \bs{x})v^{i}v^{j}-\frac{4m}{r^3}(\bs{\hat{N}}\cdot \bs{x})\hat{n}^{(i}x^{j)}-\frac{30\dot{r}^2}{r^2}(\bs{\hat{N}}\cdot \bs{x})\hat{n}^{(i}x^{j)}+
\right.
\nonumber\\[0.1cm]
&\left.
+\frac{6v^2}{r^2}(\bs{\hat{N}}\cdot \bs{x})\hat{n}^{(i}x^{j)}+\frac{12\dot{r}}{r^2}(\bs{\hat{N}}\cdot \bs{x})v^{(i}x^{j)}
\right)\,.
\end{align}
In Eq.~\eqref{eq:hij_BD_2PN}, the factor $(1-\delta_C)$ ensures that only the k-mouflage and Vainshtein cases are taken into account. We note that these contributions to the gravitational waveform originate from the scattering of the GW with the radiation-zone scalar field, which is itself sourced by the matter source in the near zone. The waveform $h^{ij}\e{GR}$ can be found in WW96 (Sec.~6), where the gravitational constant should be modified as $G\rightarrow  G/\phi_0$.

\subsection{Two-body problem and the effect of the near-zone scalar field} \label{sec:two_body_pb_correction}

The screened scalar field effectively modifies the metric at Newtonian and 1PN order, hence it is expected to affect the motion of the two black holes at these orders as well. Since we restrict the calculations of the scalar contribution to 1PN, we do not need higher PN orders of the two-body problem. Following the strategy in WW96 and Lang14, we need to find the relative acceleration $\bs{a}$ between the two black holes using their equations of motion. Furthermore, using the effective Lagrangian that recovers the equations of motion, we find the positions $\bs{x}_j$ and velocities $\bs{v}_j$ of black holes $j=1$ and $j=2$ with respect to the relative position $\bs{x}$ and velocity $\bs{v}$ in the centre of mass frame.

Starting with the relative acceleration, it can be found using the geodesic equation for both black holes and taking the difference between the two. Mathematically this gives 
\begin{align}
    \bs{a}&=\ddot{\bs{x}}_1-\ddot{\bs{x}}_2\, ,
    \\
    \ddot{x}^i_{1,2}&=-\Gamma^i_{\alpha\beta}\,\dot{x}^\alpha_{1,2}\,\dot{x}^\beta_{1,2}+\Gamma^0_{\alpha\beta}\,\dot{x}^\alpha_{1,2}\,\dot{x}^\beta_{1,2}\, \dot{x}^i_{1,2}\, ,\label{eq:geodesic_equation}
\end{align}
where $\Gamma^\mu_{\alpha\beta}$ are the Christoffel symbols. We recall that the dot notation represents the derivative with respect to coordinate time $t$, such that the geodesic equation is expressed using coordinate time, rather than proper time, thus taking a different form than the usual  geodesic equation. Rewriting the equations of motion at Newtonian and PN order leads to the result
\begin{align}
    \bs{a}=&\,\bs{a}_\textrm{N}^\0+\alpha^q\bs{a}_\textrm{N}^\q+\bs{a}_\textrm{PN}^\0+\alpha^q\bs{a}_\textrm{PN}^\q\, ,
    \\
    \bs{a}_\textrm{N}^\q=& -\frac{1}{2}\left(\psi_1'(r)+\psi_2'(r)\right)\bs{\hat{n}}\, ,\label{eq:acceleration_newtonian}
    \\[0.1cm]
    \bs{a}_\textrm{PN}^\q=&\,\frac{1}{r^2}(m_2\psi_1+m_1\psi_2+3m_1\psi_1+3m_2\psi_2)\bs{\hat{n}}+\dot{r}(\vb*{\gamma}_1'-\vb*{\gamma}_2')+\left(\Phi_{Sc,1}'+\Phi_{Sc,2}'\right)\bs{\hat{n}}\,+
    \nonumber \label{eq:a_N^q}\\
    &+\frac{1}{r}\left(m_2\psi_1'+m_1\psi_2'-3m_1\psi_1'-3m_2\psi_2'\right)\bs{\hat{n}}+\left(m_2^2\psi_1'+m_1^2\psi_2'\right)\left(\frac{v^2}{2m^2}\,\bs{\hat{n}}-\frac{
    1}{2m^2}\dot{r}\,\bs{v}\right)-
    \nonumber\\
    &-\frac{1+\mu}{2m}\left(m_2\psi_1'+m_1\psi_2'\right)\dot{r}\bs{v}+\frac{m_2}{2m}\left(\dot{r}\,\vb*{\gamma}_1'-(\bs{v}\cdot\vb*{\gamma}_1')\bs{\hat{n}}\right)-\frac{m_1}{2m}\left(\dot{r}\,\vb*{\gamma}_2'-(\bs{v}\cdot\vb*{\gamma}_2')\bs{\hat{n}}\right)-
    \nonumber\\[0.1cm]
    &-\frac{1}{m^2}\left(m_2^2\psi_1'+m_1^2\psi_2'\right)\dot{r}\bs{v}-\frac{1}{2\pi r^5}\left(m_1\psi_1+m_2\psi_2-\frac{r}{4}m_1\psi_1'+\frac{r}{4}m_2\psi_2'\right)\bs{\hat{n}}+
    \nonumber\\
    &+\frac{1}{8 m^2 \pi r^4}\bigg[\left(m_2^2\psi_1+m_1^2\psi_2\right)\left(6\,\dot{r}\,\bs{v}+9\,\dot{r}^2\,\bs{\hat{n}}\right)-3\left(m_2^2\psi_1'+m_1^2\psi_2'\right)r\,\dot{r}^2\,\bs{\hat{n}}+
    \nonumber\\[0.1cm]
    &\qquad\qquad\quad+16\pi\left(m_1^2\psi_2+m_2^2\psi_1\right)v^2\,\bs{\hat{n}}-4\pi\left(m_2^2\psi_1'+m_1^2\psi_2'\right)r\,v^2\,\bs{\hat{n}}\bigg]
    \, ,
\end{align}
and $\bs{a}_\textrm{N}^\0$, $\bs{a}_\textrm{PN}^\0$ are the Newtonian and post-Newtonian accelerations from GR and can be found in Appendix~\ref{app:CalculatingTheGravitationalWaveform}. Primes denote derivatives with respect to $r$. When calculating the Christoffel symbols, we made use of the Newtonian relations $\bs{a}_1=-m_1/r^2\, \bs{\hat{n}}$ and $\bs{a}_2=m_2/r^2\, \bs{\hat{n}}$ as well as
\begin{align}
    \pdv{\bs{v}_1}{x^j_2}=\frac{1}{2r}\bs{v}_1 \hat{n}^j\quad\textrm{and}\quad \pdv{\bs{v}_2}{x^j_1}=-\frac{1}{2r}\bs{v}_2 \hat{n}^j
\end{align}
from $\abs*{\bs{v}_i}\propto 1/\sqrt{r}$.

To find the velocities $\bs{v}_j$, $j=1,2$ with respect to the relative velocity, we can make use of the vanishing total momentum $\bs{P}=0$ in the centre of mass frame. The latter is derived from the total Lagrangian, that is deduced using the equations of motion \eqref{eq:geodesic_equation}. Since we are only interested by the waveform at 1PN, it is sufficient to find the velocities at Newtonian order. The Lagrangian at this order can be written as
\begin{align}
\mathcal{L}_\textrm{N}\equiv\mathcal{L}_\textrm{N}^\0+\alpha^q\mathcal{L}_\textrm{N}^\q =\frac{m_1}{2}\bs{v}_1^2+\frac{m_2}{2}\bs{v}_2^2+\frac{m_1m_2}{r}-\alpha^q\frac{1}{2}(\psi_1(r)+\psi_2(r)),
\end{align}
where $\mathcal{L}_\textrm{N}^\q$ can easily be found using Eq.~\eqref{eq:acceleration_newtonian}. Therefore, since we take the derivative of the Lagrangian with respect to the velocities to find the momentum and $\mathcal{L}_\textrm{N}^\q$ only depends on the position of both objects, we recover Newtonian gravity, namely
\begin{align}
    \bs{P}_\textrm{N}\equiv\pdv{\mathcal{L}_\textrm{N}}{\bs{v}_1}+\pdv{\mathcal{L}_\textrm{N}}{\bs{v}_2}= \bs{P}_\textrm{N}^\0\,.
\end{align}
Hence the velocities with respect to the relative velocity at the required order simply becomes
\begin{align}
    \bs{v}_1=\frac{m_2}{m}\bs{v}
    \quad\textrm{and}\quad
    \bs{v}_2 =-\frac{m_1}{m}\bs{v}\,.\label{eq:vi_wrt_v}
\end{align}

The individual positions $\bs{x}_j$ with respect to the relative position are found by integrating the velocities \eqref{eq:vi_wrt_v} and we simply find 
\begin{align}
    \bs{x}_1=\frac{m_2}{m}\bs{x}
    \quad\textrm{and}\quad
    \bs{x}_2 =-\frac{m_1}{m}\bs{x}\,.
\end{align}

\subsection{near-zone corrections at 1PN beyond quadrupole formula} \label{sec:waveform_1PN_correction}

The equations of motion for the binary system are already modified at Newtonian order through $\bs{a}_\textrm{N}^{(q)}\neq 0$ in Eq.~\eqref{eq:a_N^q}. Therefore we expect the quadrupole formula at 0PN to be affected. Nonetheless, the effect of this extra scalar force is suppressed by the small parameter $\alpha^q$ with respect to the GR contribution. Using the correction term~\eqref{eq:correction_EW_moment} to the Epstein-Wagoner quadrupole moment and the 
results from Sec.~\ref{sec:two_body_pb_correction}, we can find the gravitational waveform up to 1PN beyond the quadrupole approximation of a binary black hole inspiral affected by a screened scalar field. The part of the waveform $h^{ij}_\textrm{GR}$ described by GR  can be found in WW96 (Sec.~6), where the gravitational constant is replaced by $G\rightarrow G/\phi_0$ and $h^{ij}_\textrm{BD}$ enters only at higher PN order. We calculate the extra scalar contributions $\alpha^q h^{(q)ij}$ for which the result to 1PN is 
\begin{align}
    h^{ij}=&\ h_\textrm{GR}^{ij}+\alpha^q h^{(q)ij}+\mathcal{O}(\alpha^{2q})\,,
    \\[0.2cm]
    h^{(q)ij}=&\ \frac{\mu}{2R}\left[Q^{(q)ij}+P^{0.5}Q^{(q)ij}+P^{1}Q^{(q)ij}+...\right]\,,
\end{align}
where
\begin{align}
    Q^{(q)ij}=&\ -r\left(\psi_1'+\psi_2'\right)\hat{n}^i \hat{n}^j\, ,\label{eq:correction_Qqij}
    \\[0.4cm]
    P^{0.5}Q^{(q)ij}=&\ \frac{\delta m\, r}{2m}\bigg[ 
    (\psi_1'+\psi_2')\left((\bs{v}\cdot \bs{\hat{N}}-\dot{r}\,\bs{\hat{n}}\cdot \bs{\hat{N}})\, \hat{n}^i\hat{n}^j-6(\bs{\hat{n}}\cdot \bs{\hat{N}})v^{(i}\hat{n}^{j)}\right)+
    \nonumber\\
    &\qquad+(\bs{\hat{n}}\cdot\bs{\hat{N}})\,r\dot{r}(\psi_1''+\psi_2'')\hat{n}^i\hat{n}^j\bigg]\, ,
    \\[0.4cm]
    P^1 Q^{(q)ij}=&\ 2r\, a_\textrm{PN}^{(q)(i}\hat{n}^{j)}+
    \nonumber\\[0.1cm]
    &+\frac{\hat{n}^i\hat{n}^j}{12}\left[\left(16-42\eta+(2-6\eta)(\bs{\hat{n}}\cdot \bs{\hat{N}})^2\right)m(\psi_1'+\psi_2')-6(1-3\eta)r^2\dot{r}^2(\psi_1''+\psi_2'')+
    \right.
    \nonumber\\[0.1cm]
    &\left.
    \qquad+6(1-3\eta)(\psi_1'+\psi_2')\left(r\dot{r}(\bs{\hat{N}}\cdot\bs{v})-2r v^2+4(\bs{\hat{n}}\cdot \bs{\hat{N}})m\right)+
    \right.
    \nonumber\\[0.1cm]
    &\left.
    \qquad+\frac{6}{m}\bigg(\big(m+r(\dot{r}^2-v^2)\big)(m_1\psi_1'+m_2\psi_2')-r^2\dot{r}^2(m_1\psi_1''+m_2\psi_2'')\bigg)
    \right]
    \nonumber\\[0.2cm]
    &+r\, \hat{n}^{(i}v^{j)}\left[
    (3\eta-1)(\psi_1'+\psi_2')\bigg(2\dot{r}+(\bs{\hat{n}}\cdot \bs{\hat{N}})(4\bs{\hat{N}}\cdot\bs{v}-\dot{r}\,\bs{\hat{n}}\cdot\bs{\hat{N}})\bigg)-
    \right.
    \nonumber\\[0.1cm]
    &\left.
    \qquad -\frac{2\dot{r}}{m}(m_1\psi_1'+m_2\psi_2')+(3\eta-1)(\bs{\hat{n}}\cdot\bs{\hat{N}})r\dot{r}(\psi_1''+\psi_2'')
    \right]-
    \nonumber\\[0.2cm]
    &-\frac{v^iv^j}{2m}\left[
    2(m_1\psi_1+m_2\psi_2)+4(1-3\eta)mr(\bs{\hat{n}}\cdot\bs{\hat{N}})^2(\psi_1'+\psi_2')
    \right].
\end{align}
The corrections at $0$PN and $0.5$PN are only due to the modification of the equations of motion of the binary system at Newtonian order, while the scalar contributions at $1$PN originate from the extra scalar force at $0$PN and $1$PN as well as from the modification of the Epstein-Wagoner quadrupole moment.

\subsection{Example: Galileon cosmology}\label{sec:example_galileon_cosmology}

With the leading corrections to GR from a scalar field in the fully screened case and in the case where near-zone effects are considered at hand, we are ready to give an order of magnitude estimate of those. As mentioned earlier, the near-zone scalar field affects the waveform at the quadrupole order, but the modification is suppressed with $\alpha^q$.
Some of the most frequently studied models exhibiting the Vainshtein mechanism are the Galileons~\cite{Nicolis:2008in}. We focus on the cubic Galileon model described by the action
\begin{align}
    S[\phi,g]=\frac{M_p^2}{2}\int\dd^4x\left(\phi R+2\frac{\omega}{\phi}X-\frac{\alpha}{\phi^2}X\Box\phi\right) \,,
\end{align}
where $\alpha$ and $\omega$ are real parameters. To have an impact on late-time cosmology and contribute to cosmic acceleration (cf.~\cite{Lombriser:2016yzn}), a possibility is to have $\alpha\sim H_0^{-2}$, when\footnote{The cubic interaction gives an additional contribution to the kinetic term, avoiding ghost instabilities \cite{Silva:2009km}.} $\omega\sim -10\ $  \cite{Silva:2009km}. 
This is the same order of magnitude as in Ref.~\cite{Chow:2009fm}, where the cubic Galileon associated to the DGP model~\cite{Dvali:2000hr,Lombriser:2009xg} is considered. We therefore use these orders of magnitude for our comparison. In  the following, we adopt units such that $G/\phi_0=c=\hbar=1$.

We start by calculating the near-zone contribution of the scalar field. Following the scaling method \cite{McManus:2016kxu}, we find that the unique value for $q$ in the case of the Vainshtein mechanism is $q=-1/2$ and hence, the damping parameter is of the order $\alpha^q\sim H_0$. The field equation for the PN scalar field $\psi$ is found using Eq.~\eqref{eq:scalar_equation_leading_PN} and is written as
\begin{align}
    4 \pi \sums_{A=1}^2 m_A\, \delta^3(\bs{x}-\bs{x}_A)=\bigg((\bs{\nabla}^2\psi)^2-\nabla_i\nabla_j\psi\nabla^i\nabla^j\psi\bigg)\,.\label{eq:scalar_equation_cubic_Galileon}
\end{align}
To solve this equation for the binary case, we follow the strategy presented in Ref.~\cite{Hiramatsu:2012xj} and we define $\psi(\bs{x})=\psi_1(\bs{x})+\psi_2(\bs{x})+\psi_\Delta(\bs{x})$, where $\psi_{1,2}$ is the scalar field generated by the body $1,2$ respectively and $\psi_\Delta$ includes  the non-linearities. $\psi'_{1,2}(r)$ are obtained from Eq.~\eqref{eq:scalar_equation_cubic_Galileon} assuming spherical symmetry and we solve them for each body separately. Then, we find $\psi'_\Delta(r)$ by substituting the solutions $\psi'_{1,2}(r)$ into Eq.~\eqref{eq:scalar_equation_cubic_Galileon} and we assume cylindrical symmetry around the axis perpendicular to the orbital plane for the whole system. 
We require the solution to vanish asymptotically, setting an integration constant to zero, and get 
\begin{align}
    \psi'(r)=\sqrt{\frac{m_1+m_2}{2 r}}\,,\label{eq:solution_scalar_Galileon}
\end{align}
for which the PN order is $\mathcal{O}(\epsilon^{1/2})$, which is what we expect from Eq.~\eqref{eq:scalar_equation_cubic_Galileon}.
The contribution of this extra degree of freedom on the gravitational waveform is given by Eq.~\eqref{eq:correction_Qqij} as
\begin{align}
    \alpha^q Q^{(q)ij}=-\alpha^q\, r\left(\psi'(\bs{x}_1)+\psi'(\bs{x}_2)\right)\hat{n}^i\hat{n}^j=-\alpha^q m\sqrt{\frac{2r}{\mu}}\hat{n}^i\hat{n}^j\,, \qquad \forall r< \mathcal{R}\,.\label{eq:Qqij_Galileon}
\end{align}
We recall that $\alpha^q$ also carries a PN order, in this case $\mathcal{O}(\epsilon^{1/2})$, so that the overall PN order of the correction to the quadrupole formula is the same as that of GR.

The leading contribution from the radiation zone in the fully screened case can be found directly from the waveform formula~\eqref{eq:fully_screened_P15}. We consider quasi-circular orbits, for which the relative velocity $\bs{v}$ is perpendicular to the relative position $\bs{x}$ at Newtonian order and $\dot{r}\simeq 0$. Furthermore, using Newtonian approximation, the norm of the relative velocity is $v\approx \sqrt{\mu/r}$. The correction to the gravitational waveform is then given by
\begin{align}
    P^{1.5}Q^{ij}_\textrm{BD}=\frac{16}{3(3+2\omega)} \frac{m^2}{r^2}\frac{\mu}{r}\hat{n}^{(i}\hat{v}^{j)}\,,\label{eq:P15Qij_Galileon}
\end{align}
where $\bs{\hat{v}}\equiv \bs{v}/||\bs{v}||$ is the unit velocity vector. This correction is comparable to the 1.5PN order terms of GR. In comparison, the GR quadrupole of the gravitational waveform may be approximated as
\begin{align}
    Q^{ij}_\textrm{GR}=2\big(v^{(i}v^{j)}-\frac{m}{r}\hat{n}^i\hat{n}^j\big)\simeq-2\frac{m}{r}\hat{n}^i\hat{n}^j\,.\label{eq:Qij_GR}
\end{align}
The relative amplitude of each contribution, Eqs.~\eqref{eq:Qqij_Galileon} and \eqref{eq:P15Qij_Galileon}, may be compared to Eq.~\eqref{eq:Qij_GR} by neglecting factors of order unity. We obtain
\begin{align}
|\alpha^q Q^{(q)ij}| & = \frac{\alpha^q}{\sqrt{\mu}} r^{3/2}|Q^{ij}\e{GR}|\,, \label{eq:ratio1}\\
|P^{1.5}Q^{ij}_\textrm{BD}| & = \frac{1}{3+2\omega} \frac{m\mu}{r^2}|Q^{ij}\e{GR}|\, .\label{eq:ratio2}
\end{align}
By comparing $\alpha^q Q^{(q)ij}$ and $Q^{ij}_\textrm{GR}$, one can deduce that for a binary system with higher masses and higher orbital period (i.e., higher $r$), the effect of the screened scalar field is
maximised compared to the GR quadrupole formula. The scalar contribution is also strengthened when the difference between the two masses is high, minimising $\mu$ in the denominator of Eq.~\eqref{eq:ratio1}. In contrast, the effect of the Brans-Dicke scalar field in $P^{1.5}Q^{ij}_\textrm{BD}$ is amplified compared to that of GR for smaller and heavier systems, but attenuated for asymmetric masses.

Hence, we consider two different systems for the comparison; the binary black holes that produced the first detected GW \cite{Abbott:2016blz} and the inspiral of a stellar black hole with $m_1=10\, m_\odot$ around Sagittarius A$^*$ with $m_2=4\cdot 10^6\, m_\odot$, as may be observed in the close future by LISA~\cite{Barausse:2020rsu,Audley:2017drz}. The maximal amplitudes for GW150914 are
\begin{align}
    \alpha^q Q^{(q)ij}&\sim 4\times 10^{-20}\,Q^{ij}_\textrm{GR}\, ,
    \\[0.1cm]
   P^{1.5}Q^{ij}_\textrm{BD}&\sim 3\times 10^{-3}\,Q^{ij}_\textrm{GR}
\end{align}
for  the frequency range $35$-$200$~Hz. In particular, Brans-Dicke effects are stronger at higher frequencies, while near-zone contributions are amplified at lower ones. In the case of a stellar black hole around Sagittarius A$^*$, The observable frequency range of LISA is expected to be between $0.1$ mHz and $1$ Hz, which leads to
\begin{align}
    \alpha^q Q^{(q)ij}&\sim 5\times 10^{-12}\,Q^{ij}_\textrm{GR}\,,
    \\[0.1cm]
   P^{1.5}Q^{ij}_\textrm{BD}&\sim 5\times 10^{-9}\,Q^{ij}_\textrm{GR}\,.
\end{align}
In the second case, the radiation-zone contribution is weakened due to the mass asymmetry, but in the case of a supermassive black hole merger with objects of comparable masses, the Brans-Dicke effects would be of the order $(10^{-3}-10^{-2}) Q^{ij}_\textrm{GR}$. The long-range effect of the scalar field has therefore a stronger impact on the gravitational waveform with the 1.5PN term lying about one order of magnitude below LISA sensitivity~\cite{Belgacem_2019}. The near-zone correction on the waveform of the screened scalar field can be considered negligible in all cases of interest.

\section{Conclusions}
\label{sec:Conclusions}

In this work, we have investigated the impact of screening mechanisms on the GW waveform from compact sources. 
Our results are explicitly applicable to the three most prominent screening mechanisms: the Vainshtein, k-mouflage and chameleon effects. 
For our calculations we employed the modified field equations in their relaxed form and solved them with retarded integrals over the manifold. 
We first considered the case where the system is fully screened, i.e., described by GR, in the near zone and Brans-Dicke theory in the radiation zone.
The radius distinguishing these zones is identified as the screening radius of the underlying gravity theory.
It was found that the boundary terms exactly cancel between the near and radiation zone and the screening radius does not enter the waveform.
The deviations from the GR waveform that this model induces appear at 1.5PN beyond the quadrupole formula.

Further, we found the leading corrections in the near zone to the fully screened results induced by the scalar field. 
In this case, we derived the leading corrections in the stress-energy tensor modifying the Epstein-Wagoner moments at 1PN.
The near-zone metric was also found to be modified at Newtonian and post-Newtonian orders, leading to 0PN, 0.5PN and 1PN corrections beyond the quadrupole formula from the scalar force. 
Although these contributions have low PN order, they are suppressed by 
the nature of screening, which consequently makes these deviations subdominant to the Brans-Dicke corrections found at 1.5PN from the unscreened radiation zone. Importantly, in Chameleon models the waveform is unaffected for a black hole source by the scalar field, but the effect is present for other compact objects, such as neutron stars.

Finally, we applied our results to a Galileon cosmology toy model and estimated the amplitude of the deviation from GR for a black hole binary. 
We found that in the most optimistic scenarios the waveform amplitude, and hence luminosity distance, deviates from its GR counterpart by one part in $10^{11}$ from the modification of the near-zone metric by the screened scalar field and one part in $10^{2}$ from the radiation-zone Brans-Dicke field. 
While the near-zone corrections are not realistically measurable in the foreseeable future, radiation-zone contributions might be observable with future detectors.
In particular, LISA and the Einstein telescope are forecasted to have an order $10^{-2}$ error on the luminosity distance~\cite{Belgacem_2019,Maggiore:2019uih}. 
Hence the effect of a screened Horndeski theory on the waveform may be detectable at the 1-$\sigma$ level in the near future. 
The predicted radiation-zone contributions depend on the Brans-Dicke parameter and we expect future GW measurements to set new constraints on this parameter.

While the effect of screening on the waveform is small, it may still play an important role in multi-band gravitational wave detection \cite{Barausse:2016eii}.
The LISA mission will detect binaries several years before merger~\cite{Barausse:2020rsu,Audley:2017drz,Gerosa:2019dbe}.
Consequently, as the binary evolves, it exits the LISA sensitivity band only to enter ground-based detectors sensitivity bands a few years to decades later.
The cumulative effect of the modified waveform energy emission would accumulate over this time causing a dephasing of the screened signal and one predicted by GR. 
Further analysis on the energy loss due to the modified GW and the suppressed scalar wave would allow one to predict a different phase and time of merger in comparison to GR, a potential test of any given screening mechanism despite the difference in the signal itself being too small to detect directly in ground-based detectors~\cite{Gnocchi_2019}. In addition, recent relativistic numerical analysis have shown that kinetic screening might not play a role in strong field regime leaving the scalar contributions as strong as in Brans-Dicke theories \cite{Bezares:2021yek,Bezares:2021dma}.

In the analysis presented, we made several simplifying assumptions. 
First, we omitted the contribution of the scalar waves on the detectors, which may play a significant role. 
Even though such waves are screened in the emission and detection regions weakening their effects, future work should be devoted to their study.
We made the standard assumption that the orbit was circularised for late inspiral calculations.
However, even if the system were elliptical, one would expect that the modification to the waveform caused by dropping this assumption remains small since the system still lies within the screened region. 
In this work, we restricted our final calculations to the case of black hole binaries and one may wonder how the results would change for neutron star binaries or mixed systems. 
For the k-mouflage and Vainshtein mechanisms, one can argue that the field equation governing the behaviour of the near-zone scalar field satisfies Gauss's law, thus the composition of the objects does not affect their motion, only their mass does. 
This is not the case for the chameleon mechanism, where the sensitivity to the scalar field varies with the compactness. 
For theories with such screening, one would need to include near-zone effects of the scalar field in the case of neutron stars, but we expect these contributions to be strongly negligible compared to the Brans-Dicke corrections from the radiation zone.
Finally we have not considered the contributions originating from the intrinsic spin of each object.
In GR, these contributions modify the Epstein-Wagoner moments and enter in the waveform formula at 1PN beyond quadrupole order~\cite{Will:1996zj}. 
Therefore, we expect that, in the fully screened scenario treated in Sec.~\ref{sec:waveform_fully_screened}, only GR contributions affect the waveform (see Appendix~F of WW96). 
When calculating the leading corrections from the near-zone scalar field in Sec.~\ref{sec:corrections_to_waveform_from_scalar_field}, one could expect new spin contributions involving the scalar field entering at 1PN, but these terms would also be strongly suppressed by the screening mechanism.

In conclusion, this work supports the use of GW emission processes abiding by GR in the study of screened modified gravity theories. 
Nonetheless, we raised the possibility to probe the effect of a screened scalar field on GWs with the sensitivity of our future detectors. 
Even though we have only investigated cubic Galileon interactions as a specific case, this study offers an efficient method to check for a wide range of Horndeski theories whether scalar field effects can be neglected in the GW emissions of the respective models.

\section*{Acknowledgments}

The authors would like to thank Jonathan Gair and Jorge Pe\~narrubia for useful discussions at an early stage of this work in 2016/17.
C.R. was supported by the University of Grenoble-Alpes.
R.M. acknowledges support from the STFC Consolidated grant ST/T000732/1.
C.D.~and L.L.~were supported by a Swiss National Science Foundation (SNSF) Professorship grant (No.~170547).
L.L. also acknowledges the support by a SNSF Advanced Postdoc.Mobility Fellowship (No.~161058) during an early stage of this work.
Please contact the authors for access to research materials.

\appendix

\section{Two-body problem in General Relativity}
\label{app:CalculatingTheGravitationalWaveform}

For completeness we shall provide here the details of the calculation of the gravitational waveform from a binary system with masses $m_{1,2}$ and positions $\bs{x}_{1,2}$ conducted in Sec.~\ref{sec:fully_screened_binary}. In performing this calculation we change coordinates to place the centre of mass of the system at the origin. 
This is a common trick in classical physics where the mass of each body is only its rest mass.
However, we consider PN corrections to the system, and so the respective mass of an object is also changed by its potential and kinetic energy. 
Using the corrected mass changes the centre of mass and so its definition changes accordingly.

The centre of mass is found to be 
\begin{align}
\bs{X} =&\ m^{-1}(m_1 \bs{x}_1 + m_2 \bs{x}_2) + \bs{f}^{(1)} + \mathcal{O}(\epsilon^2)\,,
\\
\bs{f}^\1=&\ -\frac{1}{2}\eta\frac{\delta m}{m}\left(v^2-\frac{m}{r}\right)(\bs{x}_1 - \bs{x}_2)\,,
\end{align}
where $m = m_1 + m_2$, $\mu = m_1m_2/m$, $\eta = \mu/m$, $\delta m = m_1 - m_2$ and $r = |\bs{x}_1 - \bs{x}_2|$. 
One may worry that as we are calculating the metric to $\mathcal{O}(\epsilon^4)$, we may need the centre of mass up to $\mathcal{O}(\epsilon^2)$, but the only place
the correction enters is in the leading-order quadrupole term, where it cancels out exactly.

We wish to describe the system in terms of the relative separation $\bs{x} = \bs{x}_1 - \bs{x}_2$ and its derivatives.
In the centre of mass coordinate system, $\bs{X}=0$, we find
\begin{subequations}
\begin{align}
\bs{x}_1 =& (m_2/m) \bs{x} - \vb*{f}^\1 + \mathcal{O}(\epsilon^2)\,, \\
\bs{x}_2 =& -(m_1/m) \bs{x} - \vb*{f}^\1 + \mathcal{O}(\epsilon^2)\,.
\end{align}
\end{subequations}

It will be useful to define the system in terms of Newtonian objects in the final expression.
So we define the Newtonian angular momentum, $\bs{L} = \mu \bs{x}\times\bs{v}$ and unit normal to the orbital plane $\bs{\hat{\lambda}} = \bs{L}\times\bs{\hat{n}}$.

When computing the time derivatives of the centre of mass coordinates, we will make use of the equations of motion for the black hole binary.
As the near zone is described by GR, the equations of motion can be taken readily from sources such as Ref.~\cite{BlanchetReview}.
The accelerations due to the gravitational forces to first PN order are found to be
\begin{subequations}
\begin{align}
\label{eq:PNAcceliration}
\bs{a} =&\ \bs{a}_\textrm{N} + \bs{a}_\textrm{PN} + \mathcal{O}(\epsilon^2)\,, \\
\bs{a}_\textrm{N} =& -\frac{m}{r^2}\bs{\hat{n}}\,,\\
\bs{a}_\textrm{PN} =& -\frac{m}{r^2}(A_\textrm{PN} \bs{\hat{n}}+ B_\textrm{PN} \dot{r}\bs{v})\,,
\end{align}
\end{subequations}
where for convenience we define
\begin{subequations}
\begin{align}
A_\textrm{PN} =& -2(2+\eta)\frac{m}{r} + (1 + 3 \eta)v^2 - \frac{3}{2}\eta \dot{r}^2\,, \\
B_\textrm{PN} =& -2(2-\eta)\,.
\end{align}
\end{subequations}
Derivatives of the equations of motion are also needed.
In order to calculate $\dot{a}^{j}$ and $\ddot{a}^{j}$, we make use of the identities
\begin{subequations}
\begin{align}
\bs{\dot{\hat{n}}} =&\ \frac{\bs{v}}{r} - \frac{\dot{r}\bs{n}}{r}\,,\\
\ddot{r} =&\ \ddot{r}_\textrm{N} + \ddot{r}_\textrm{PN}\,,\\
\ddot{r}_\textrm{N} =&\ - \frac{m}{r^2} + \frac{v^2}{r} - \frac{\dot{r}^2}{r}\,,\\
\ddot{r}_\textrm{PN} =&\ - \frac{m}{r^2} ( \alpha \frac{m}{r} + \beta v^2 + (\gamma + \sigma)\dot{r}^2 ) \,,
\end{align}
\end{subequations}
which give $\dot{a}^{j}$ to the required order as
\begin{subequations}
\label{eq:dota}
\begin{align}
\dot{a}^{i} =&\ (\dot{a}_\textrm{N})_\textrm{N}^{i} + (\dot{a}_\textrm{PN})_\textrm{N}^{i}\,,\\
(\dot{a}_\textrm{N})_\textrm{N}^{i} =&\  \frac{2m\dot{r}}{r^3}\hat{n}^i - \frac{m}{r^3}v^i\,,\\
(\dot{a}_\textrm{PN})_\textrm{N}^{i} =&\ -18 \frac{m^2\dot{r}}{r^4}\hat{n}^i +4\eta \frac{m^2}{r^4}v^i + (3+12\eta)\frac{m\dot{r}v^2}{r^3}\hat{n}^i - \frac{15}{2}\eta\frac{m\dot{r}^3}{r^3}\hat{n}^i \nonumber \\ 
&-(12- \frac{15}{2}\eta)\frac{m\dot{r}^2}{r^3}v^i + (3-5\eta)\frac{mv^2}{r^3}v^i
\end{align}
\end{subequations}
and $\ddot{a}^{j}$ to the required order as
\begin{subequations}
\label{eq:ddota}
\begin{align}
\ddot{a}^{i} =&\ (\ddot{a}_\textrm{N}^i)_\textrm{N} + (\ddot{a}_\textrm{N}^i)_\textrm{PN} + (\ddot{a}_\textrm{PN}^i)_\textrm{N}\,,\\
(\ddot{a}_\textrm{N}^i)_\textrm{N} =&\ -2 \frac{m^2}{r^5}\hat{n}^i -15 \frac{m\dot{r}^2}{r^4}\hat{n}^i + 3 \frac{m v^2}{r^4}\hat{n}^i + 6\frac{m \dot{r}}{r^4}v^i  \,,\\
(\ddot{a}_\textrm{N}^i)_\textrm{PN} =&\ (8+4\eta)\frac{m^3}{r^6}\hat{n}^i - (2+6\eta)\frac{m^2v^2}{r^5}\hat{n}^i + (12-3\eta)\frac{m^2\dot{r}^2}{r^5}\hat{n}^i \nonumber \\ 
&- (4-2\eta)\frac{m^2 \dot{r}}{r^5} v^i \,,\\
(\ddot{a}_\textrm{PN}^i)_\textrm{N} =&\  -36\eta\frac{\dot{r} m^2}{r^5}v^i - (30-42\eta)\frac{m\dot{r}v^2}{r^4}v^i - (60-45\eta)\frac{m\dot{r}^3}{r^4}v^i \nonumber \\
&+ (126-57\eta)\frac{m^2\dot{r}^2}{r^5}\hat{n}^i - (15+165/2\eta)\frac{mv^2\dot{r}^2}{r^4}\hat{n}^i + 102/2\eta\frac{m\dot{r}^4}{r^4}\hat{n}^i \nonumber \\
&+ (22+2\eta)\frac{m^3}{r^6}\hat{n}^i  + (8-10\eta)\frac{m^2v^2}{r^5}\hat{n}^i + (3+15\eta)\frac{mv^4}{r^4}\hat{n}^i\,.
\end{align}
\end{subequations}

\bibliographystyle{JHEP} 

\bibliography{main}

\end{document}